\documentclass[twocolumn]{aastex63}
\usepackage{amsmath}
\usepackage{romannum}

\received{June 1, 2019}
\revised{January 10, 2019}
\accepted{\today}
\submitjournal{ApJ}

\shorttitle{D4 census}
\shortauthors{Shi et al.}

\begin{document}

\title{Accelerated galaxy growth and environmental quenching in a protocluster at z=3.24 }

\correspondingauthor{Ke Shi}
\email{keshi@xmu.edu.cn}

\author{Ke Shi}

\affiliation{Department of Astronomy, Xiamen University, Xiamen, Fujian 361005, China}

\author{Jun Toshikawa}

\affiliation{Institute for Cosmic Ray Research, The University of Tokyo, Kashiwa, Chiba 277-8582, Japan}

\affiliation{Department of Physics, University of Bath, Claverton Down, Bath, BA2 7AY, UK}

\author{Kyoung-Soo Lee}

\affiliation{Department of Physics and Astronomy, Purdue University, 525 Northwestern Avenue, West Lafayette, IN 47907, USA}

\author{Tao Wang}
\affiliation{Institute of Astronomy, University of Tokyo, 2-21-1 Osawa, Mitaka, Tokyo 181-0015, Japan}

\author{Zheng Cai}

\affiliation{Department of Astronomy, Tsinghua University, Beijing 100084, China}

\author{Taotao Fang}
\affiliation{Department of Astronomy, Xiamen University, Xiamen, Fujian 361005, China}

\begin{abstract}
We present a multiwavelength study of galaxies around D4UD01, a spectroscopically confirmed protocluster at $z=3.24$ to investigate environmental trends. 450 galaxies are selected based on $K_S$ band detection with photometric redshifts (photo-$z$) at $3.0<z<3.4$, among which $\sim12$\% are classified as quiescent galaxies. The quiescent galaxies are among the most massive and reddest ones in the entire sample. We identify a large photo-$z$ galaxy overdensity in the field, which lies close to the previously spectroscopically confirmed sources of the protocluster. We find that the quiescent galaxies are largely concentrated in the overdense protocluster region with a higher quiescent fraction, showing a sign of environmental quenching. Galaxies in the protocluster are forming faster than the field counterparts as seen in the stellar mass function, suggesting early and accelerated mass assembly in the overdense regions. Although weak evidence of suppressed star-formation is found in the protocluster, the statistics are not significant enough to draw a definite conclusion. Our work shed light on how the formation of massive galaxies is affected in the dense region of a protocluster when the Universe was only 2 Gyr old.

\end{abstract}



\section{Introduction} \label{sec:intro}
It is well known that local environments have profound impacts on the formation and evolution of galaxies. At low redshift (e.g., $z\lesssim1$), clusters contain a higher fraction of red massive ellipticals while young late-type galaxies are mostly found in the low-density field. This `morphology-density' relation \citep{Dressler80,Dressler97, Goto03, Kauffmann04, Postman05} observed out to $z\sim1$ also implies that cluster galaxies must have experienced an early growth at higher redshift. Indeed, stellar population studies of cluster galaxies suggest that they generally experienced a short and intense star-formation and quenched very quickly afterwards \citep[e.g.,][]{Stanford98,Thomas05,Snyder12,Mart18}.

In order to further investigate the environmental impacts on galaxy formation and understand the detailed quenching mechanism of cluster galaxies, we need to directly observe the progenitors of clusters (`protoclusters') and study their galaxy constituents at high redshift. Many studies have shown that star formation activities in dense environments are enhanced relative to the field at high redshift \citep[e.g.,][]{Elbaz07,Cooper08,Tran10,Koyama13,Alberts14,Cai17,Shimakawa18,Lemaux20}, suggesting protocluster galaxies may have undergone accelerated mass assembly than the field counterparts. However, there are also some of protoclusters where no such differences are seen as compared to the field \citep[e.g.,][]{Overzier08,Toshikawa14,Cucciati14,Shi191}. This discrepancy between different studies may rise from different galaxy populations they use or the lack of statistics due to small sample size \citep{Cucciati14}, or because of different evolutionary stages in which these protoclusters are observed \citep{Overzier08,Toshikawa14}. 

Despite the above challenges in identifying the reversal of `star formation-density' relation in protoclusters, a growing number of studies suggest that protoclusters often host a larger fraction of massive red galaxies that have already quenched their star-formation \citep[e.g.,][]{Steidel05,Kubo13,Lemaux14,Lemaux18,Zavala19,Shi191,Ando20}. These studies suggest that the cluster red sequence observed in the local Universe \citep{Visvanathan77,Bower92,Stott09} may have already been formed in protoclusters at earlier epochs. This is further supported by the numerical simulation of \cite{Chiang17}, who proposed an ``inside-out'' galaxy growth in protoclusters from $z>10$ to $z\sim5$, when most of the star-formation and mass assembly happen in the central regions. This growth of cores is followed by an extended star-formation in the entire protocluster region at $1.5<z<5$ when the cores begin to quench and massive quiescent galaxies would be seen.

A systematic investigation of environmental impacts requires more observations of distant protoclusters. However, protoclusters are not virialized yet, and they typically extend up to 10$\arcmin$-30$\arcmin$ in the sky \citep{Chiang13,Muldrew15}, making it difficult and observationally expensive to conduct a systematic search. While several tens of protoclusters have been spectroscopically confirmed to date \citep[see e.g.,][for a summary]{Overzier16,Harikane19}, many  were identified by pre-selecting overdense regions traced by star-forming galaxies such as Lyman break galaxies (LBGs) or Lyman alpha emitters (LAEs), followed up by spectroscopic confirmation.

\cite{Toshikawa16} discovered a protocluster in the D4 field of the Canada-France-Hawaii-Telescope Legacy Survey (CFHTLS) at $z=3.24$. This protocluster, dubbed `D4UD01' hereafter, was initially identified using $u$-dropout selected LBGs at $z\sim3$. A significant surface overdensity (4.4$\sigma$) of LBGs was found in the field, implying the presence of a large structure. Follow-up spectroscopy has confirmed five galaxies at $z=3.24$ within 2 Mpc (physical) with one another. Further comparison with simulation also suggested that it will become a virialized cluster at $z=0$ \citep{Toshikawa16}. However, LBGs are star-forming galaxies that severely suffered from projection effects due to large redshift uncertainties, therefore it is  difficult to conduct a systematic study of environmental impacts in this protocluster using only LBGs. To better characterize the role of the environments, a detailed census of its galaxy constituents is needed. In this work we perform a multiwavelength study of galaxies in and around this protocluster with the help of photometric redshift, aiming to further unveil the environmental trends in this protocluster.

This paper is organized as follows. In Section~\ref{sec2} we describe the data and methods used to select the protocluster galaxy candidates.  Massive quiescent galaxies are selected in Section~\ref{sec3}. In Section~\ref{distribution} we investigate the spatial distributions of galaxies in the field and identify an overdensity which is defined as the protocluster region. The sky distribution of quiescent galaxies is also studied to seek possible environmental trends. We further compare the GSMFs and star-formation rates of protocluster galaxies with the field counterparts in Section \ref{dis}. We summarize our results in Section~\ref{sum}. Throughout this paper we use the WMAP9 cosmology ($\Omega_M=0.29, \Omega_\Lambda=0.71, \sigma_8=0.83, h=0.69$) from \cite{Hinshaw13}. All magnitudes are given in the AB system \citep{Oke83}. Distance scales are given in comoving units unless noted otherwise.

\section{Data and Analysis} \label{sec2}

\subsection{Data and photometry} \label{data}
In this study, we make use of publicly available multiwavelength data including the deep optical $ugriz$ images from the CFHTLS Deep Servey \citep{Gwyn12} and the 
near-IR $JHK_S$ bands from WIRCam Deep Survey (WIRDS) \citep{Bielby12}. We also use the Spitzer-IRAC $3.6\micron$ and $4.5\micron$ data from the NMBS-II IRAC survey \citep{Annunziatella18}. The pixel scale of CFHTLS and WIRDS images is 0.186$\arcsec$ while for IRAC data it is 0.558$\arcsec$. The photometric depths of CFHTLS and WIRDS data are measured from the sky fluctuations by placing 2$\arcsec$ diameter apertures in random image positions while the depths of IRAC data are measured within 3$\arcsec$ aperture. Table \ref{table1} summarizes the data sensitivity and image quality in this paper. It is noted that the IRAC data is fairly shallow (maximum exposure time of only $\sim$1 hour for each channel) with large photometric uncertainties, and thus they have weak constraints in our analysis.

We created a multiwavelength photometric catalog as follows.
First, we smooth the WIRDS images and other CFHTLS bands data to match the largest point-spread function (PSF) of the CFHTLS $u$ band data. To do so, the radial profile of the PSF in each image is approximated by a Moffat function with the measured seeing FWHM. A noiseless convolution kernel between the low and high-resolution images is then derived using the Richardson-Lucy deconvolution algorithm \citep{Richardson:72}. Each  image is convolved with its respective kernel to match the PSF of the $u$ band data. 

The WIRDS survey does not cover the entire 1$^{\circ}$ $\times$ 1$^{\circ}$ D4 field (60\% of the field has no data, see \cite{Bielby12}), but fortunately D4UD01 is located within its coverage. For source detection in this work, we use the $K_S$ band which samples the rest-frame optical emission at $z=3.24$, enabling the measurement of stellar masses of the galaxies. We trim the $K_S$ image to contain the region receiving $>$50\% of the maximum exposure time, which result in a final area of 497 arcmin$^2$. All the other bands are also trimmed to match the $K_S$ image.

Source detection and photometric measurements in the $ugrizJHK_S$ bands are carried out by running the SExtractor software \citep{Bertin96} in dual mode on the PSF matched images with the $K_S$ image as the detection band. The SExtractor parameter MAG\_AUTO is used to estimate the total magnitude, while colors are computed from fluxes within a fixed isophotal area (i.e., FLUX\_ISO). As the images are PSF matched, aperture correction in all bands is assumed to be the difference between MAG\_AUTO and MAG\_ISO measured in the detection band. For sources not detected in certain bands, we use the 2$\sigma$ limiting magnitude to give the upper limits.

 As for the IRAC images, since the PSFs of these images are much broader ($\approx$ 1.8$\arcsec$), source blending on these images is a severe problem. In order to obtain accurate and unbiased measurement of fluxes and colors on the IRAC images, we utilize the T-PHOT software \citep{Merlin15,Merlin16}. T-PHOT performs ``template-fitting''
photometry on the low-resolution image using the information of high-resolution image and catalog. In our case, the $K_S$ band image and catalog are used as the input priors of T-PHOT while the low-resolution IRAC images are analysized to obtain corresponding photometry. It is noticed that although T-PHOT is prior-based, the derived photometry of the low-resolution image does not strongly depend on which high-resolution image we use as the input. For example, if we use the $i$ band as the input high-resolution prior to derive the 3.6$\micron$ photometry, the resultant 3.6$\micron$ magnitude differences as compared to the $K_S$ based have only an average value of $\sim0.04$. This confirms that our T-PHOT derived photometry is not significantly biased by the prior.

Finally, all photometric catalogs are combined together to make a multiwavelength catalog. In this work,  we focus on the sources with $K_S$ magnitudes smaller than 24.29 (i.e., $>$ 5$\sigma$ detection limit). In the end 20,801 sources are selected in the final catalog. 

\begin{deluxetable}{cccc}[h]
\tablecaption{Data Set \label{table1}}
\tablehead{
\colhead{Band} & \colhead{Instrument} & \colhead{Limiting magnitude\tablenotemark{$\star$}} & \colhead{FWHM} \\
\colhead{} & \colhead{} & \colhead{(5$\sigma$,AB)} & \colhead{($\arcsec$)}
}
\startdata
$u$ & MegaCam/CFHT & 27.02 & 0.90\\
$g$ & MegaCam/CFHT & 27.48 & 0.80\\
$r$ & MegaCam/CFHT & 27.11 & 0.70\\
$i$ & MegaCam/CFHT & 26.72 & 0.70\\
$z$ & MegaCam/CFHT & 25.84 & 0.70\\
$J$ & WIRCam/CFHT & 24.83 & 0.60\\
$H$ & WIRCam/CFHT & 24.33 & 0.60\\
$K_S$ & WIRCam/CFHT & 24.29 & 0.60\\
3.6 $\micron$ & IRAC/Spitzer & 22.27 & 1.86\\
4.5 $\micron$ & IRAC/Spitzer & 22.29 & 1.75\\
\enddata
\tablenotetext{\star}{ 5$\sigma$ limiting magnitude measured in a 2$\arcsec$ diameter aperture for the CFHT data, while for the Spitzer data the depths are measured in a 3$\arcsec$ aperture. 
}
\end{deluxetable}


\subsection{Photometric Redshift and Spectral Energy Distribution Fitting} \label{sedfitting}
The photometric redshift and the physical properties of each source in the catalog are derived via the spectral energy distribution (SED) fitting technique using the CIGALE software \citep{Noll09, Boquien19}. Based on an energy balance principle (the energy emitted by dust in the mid- and far-IR exactly
corresponds to the energy absorbed by dust in the UV-optical
range), CIGALE builds composite stellar population models from various single stellar population models, star formation histories, dust attenuation laws, etc. The model templates are then fitted to the observed fluxes of galaxies from far-ultraviolet to the radio domain, and photometric redshift as well as physical properties are estimated using a Bayesian analysis.

For the SED templates, we use the stellar population synthesis models of \cite{BC03} , \citet{Calzetti00} reddening law with E(B-V) values
ranging from 0 to 2 in steps of 0.1 mag, the solar metallicity,
and  \cite{Chabrier03} initial mass function. We use the delayed star formation history (SFR $\propto$ t $\times$ exp[-t/$\tau$]) with star-forming time scale $\tau$ ranging from 0.1 to 10 Gyr. Nebular emission is also included and dust emission is modeled by \cite{Dale14}. The input redshifts are set to be between 0.1 and 5.0 in steps of 0.1. 

To calibrate the photometric redshift (photo-$z$), we use a spectroscopic redshift (spec-$z$) sample obtained from the AAOmega instrument on the Anglo-Australian Telescope (AAT) targeting X-ray point sources in the D4 field \citep{Stalin10}. The sample has 1,809 spec-$z$ sources with the majority of them lying at $z<1$. We also use the 16 spectroscopic LBGs in \cite{Toshikawa16} in D4UD01. We crossmatch these spec-$z$ sources with our photo-$z$ catalog and find 191 counterparts. The precision of the photo-$z$ is measured using the normalised median absolute deviation \citep{Hoaglin83} defined as $\sigma_z=1.48\times$ median($\mid\Delta_z\mid$/(1+$z_\mathrm{spec}$)), where $\Delta_z=z_\mathrm{spec}-z_\mathrm{phot}$. This scatter measurement corresponds to the rms of a Gaussian distribution and is not affected by catastrophic outliers (i.e., objects with $\mid\Delta_z\mid$/(1+$z_\mathrm{spec}$)$>0.15$) \citep{Ilbert06,Laigle16}. For these sources, we obtain $\sigma_z=0.08$. The number of catastrophic failures take up to 10\% of all the sources.

The mean photo-$z$ error derived by CIGALE is $\Delta z \sim0.2$ in our sample, therefore we select 783 galaxies with photo-$z$ measurements of $3.0<z_\mathrm{phot}<3.4$ as potential protocluster galaxy candidates, so that the protocluster redshift ($z=3.24$) lies within the coverage of the photo-$z$ error. Only three objects among these candidates have spec-$z$ information. One ($z_{\textrm{phot}}=3.1$) is in the AAOmega sample that indicates it is a QSO at $z_{\textrm{spec}}=3.03$, and we remove this object in our candidate list. The other two ($z_{\textrm{phot}}=3.3$ and $3.0$) are the spectroscopically confirmed LBGs at $z_{\textrm{spec}}=3.24$ and $3.04$ in \cite{Toshikawa16}. The remaining 14 spec-$z$ sources in \cite{Toshikawa16} are not detected in $K_S$ band, therefore not in our photo-$z$ sample.  For the 782 sources, we remove the ones that have SExtractor parameter ``CLASS$\_$STAR'' greater than 0.9 to reduce the contamination of stars. We then visually inspect the remaining sources and remove those with potential contamination in the photometry, including those severely blended with nearby bright sources. We also discard sources that are detected in less than four bands. In the end, 450 galaxies are selected as our photo-$z$ galaxies.

We fix the best-fit photo-$z$ of the 450 galaxies and refit their SEDs using CIGALE with the same configuration to determine their physical properties such as stellar mass, star formation rate (SFR) and color excess of stellar continuum E(B-V), etc. The typical (median) error of stellar mass is $\sim0.1$ dex while for SFR it is $\sim0.3$ dex. 

For all the photo-$z$ galaxies, we also estimate their stellar mass completeness following an empirical approach \citep{Pozzetti10,Ilbert13,Laigle16}. For each  galaxy, we compute the lowest stellar mass $M_\mathrm{lim}$ it would need to be detected at the given $K_S$ magnitude limit $K_S$ $_\mathrm{lim}=24.29$:
\begin{equation*}
\mathrm{log}(M_\mathrm{lim})=\mathrm{log}(M)-0.4(K_S\,_\mathrm{lim}-K_S),
\end{equation*}
and the stellar mass completeness limit corresponds to the mass under which 90\% of the galaxies lie. The calculated mass completeness limit is $\mathrm{log}(M_\mathrm{lim})=10.8$ in our photo-$z$ sample. We also notice that only 193 (43\%) of the photo-$z$ galaxies satisfy the LBG criteria defined in \cite{Toshikawa16}. LBGs are believed to be young star-forming galaxies with typical stellar masses of $10^{10}\mathrm{M_\sun}$ \citep[e.g.,][]{Giavalisco02}. Therefore our sample includes a large fraction of
massive galaxies that are commonly missed from the UV-selected LBGs, which is helpful in studying the high-mass end of the stellar mass function.

We also consider possible contamination in our sample. The photo-$z$ galaxies lie around at $z\approx3.2$ where the $K_S$ band photometry could be potentially contaminated by the [O~{\sc iii}]$\lambda\lambda$4959,5007 nebular emission lines, which would lead to an overestimate of the stellar mass derived from SEDs. For example, \cite{Schenker13} measured the rest-frame
[O~{\sc iii}] equivalent widths (EWs) for a sample of $3.0<z<3.8$ LBGs and determined an average value of 250 \AA. At $z\approx3.2$, this leads to an overestimate of $K_S$ band continuum flux density by 0.3 magnitude. However,  \cite{Malkan17} noticed there is an anti-correlation between the stellar mass and [O~{\sc iii}] EW for LBGs at $z\sim3$: galaxies with higher stellar masses usually have smaller EWs. According to their relation, 98\% galaxies in our sample with masses $>10^{10} M_\sun$ have typical EWs less than 100 \AA, corresponding to a flux contamination smaller than 0.1 magnitude. More recently, \cite{Yuan19} further investigated the impact of including [O~{\sc iii}] nebular emission data in the SED fitting analysis using a sample of LBGs at $z\sim3.5$. They found an average discrepency of only $\sim0.1$ dex in the derived stellar mass when nebular emission data is included in the fitting. In comparison, the average stellar mass uncertainty of our photo-$z$ sample is also $\sim0.1$ dex. Thus we conclude the influence of nebular emission on the derived stellar mass is minimal and does not significantly affect the main results of this paper.

\section{Selection of Quiescent Galaxy Candidates} \label{sec3}

The presence of quiescent galaxies at high redshift can give us valuable insight into how current-day massive ellipiticals obtain their masses. One of the main focus of this paper is to identify and study evolved galaxy populations in the protocluster field. To do so, a reliable method to separate quiescent galaxies from star-forming galaxies is required.

Various methods have been developed to classify quiescent galaxy populations in the literature. Perhaps the most well-known method is using the rest-frame $U-V$ vs. $V-J$ color-color diagram ($UVJ$ diagram), where the galaxy distributions are bimodal and a color cut can be applied to separate the two populations \citep[e.g.,][]{Labbe05,Williams09, Brammer11, Muzzin13}. Other color criteria have also been proposed, such as the NUV$-r$ vs. $r-J$ \citep{Ilbert13} which can alleviate the confusion between red dusty star-forming galaxies and passive galaxies, and $J-K_S$ vs. [3.6]-[4.5] color in the observed frame that select galaxies with a strong Balmer/4000\AA\ break at $2<z<4$ \citep{Girelli19,Shi20}.

In this work, as the IRAC bands are too shallow to give strong constraints in the rest-frame near-IR (see Section \ref{data}), the above color-color criteria may not be appropriate in separating quiescent galaxies. Therefore we will take a different approach, using the 4000\AA\ break index (D4000 hereafter). The spectral break at 4000\AA\ is the strongest discontinuity in the optical spectrum of a galaxy, which is mainly caused by ionized metal lines (e.g., Ca\Romannum{2}~H and K lines) in older stellar populations. A break index D4000 is defined as the ratio of the average flux density at the wavelength of 4000-4100 \AA\ and 3850-3950 \AA\ \citep{Balogh99}. This definition of using narrow bands has its advantage of being less sensitive to dust attenuation. D4000 can be regarded as a stellar population age indicator: a larger value usually suggests an older age of the galaxy, and thus can be used as a criterion to separate young star-forming galaxies and old quiescent galaxies \citep[e.g.][]{Kauffmann03,Gallazzi05,Hathi09,Johnston15,Haines17}. Furthermore, it has been shown that D4000 is also closely correlated with specific star-formation rate (sSFR, defined as SFR/M$_\mathrm{star}$) where a larger value typically corresponds to a lower sSFR \citep{Brinchmann04}. 

\begin{figure*}[ht!]
\plotone{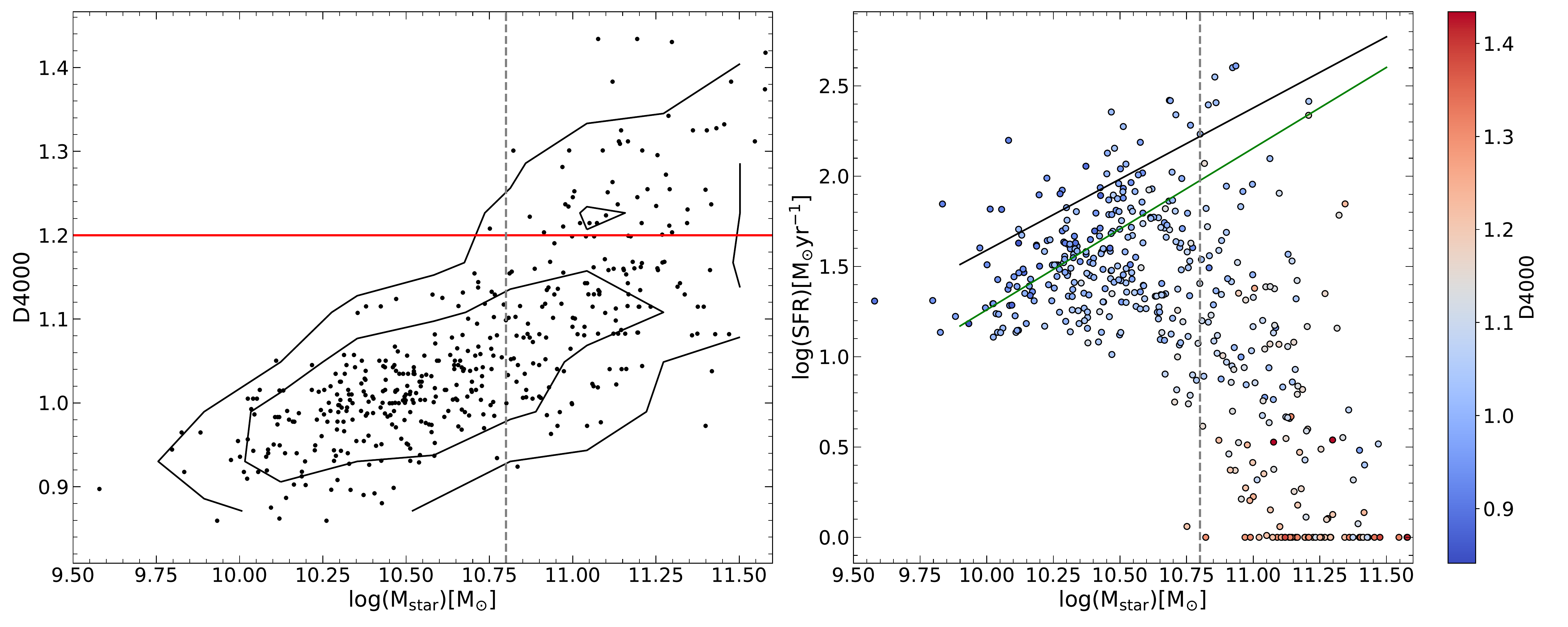}
\caption{
 {\it Left:} D4000 values of the photo-$z$ galaxies as a function of stellar mass. A bimodality can be seen in the plane from the 1$\sigma$ and 2$\sigma$ contour lines. The red line indicates our selection criterion for quiescent galaxies. The dashed vertical line denotes the 90\% mass completeness of the sample.
 {\it Right:} SFR--$\mathrm{M_{star}}$ relation for the photo-$z$ galaxies color coded by their D4000 values. The black line is the main-sequence relation calibrated from \cite{Speagle14} and the green line is that from a semi-analytic model by \cite{Dutton10}. Galaxies with SFR=0 are indicated in the log(SFR)=0 location.
}
\label{fig:D4000}
\end{figure*}

In this work, D4000 index is inferred from the best-fit template of CIGALE. The typical error of D4000 value is $\sim0.04$. In the left panel of Figure \ref{fig:D4000}, we show the D4000 distribution as a function of stellar mass. As can be seen in the figure, the distribution appears to be bimodal: most of the galaxies are located in the lower left corner whereas a fraction are concentrated in the upper right corner. This bimodality has been seen both at low redshift ($z<1$) \citep{Haines17} and at high redshift (up to $z\sim3$) \citep{Johnston15}. Based on this diagram, we apply a cut at D4000=1.2, which roughly segregate the two populations, defining galaxies above this limit to be quiescent galaxy candidates. In total, 52 galaxies fall into the quiescent galaxy catalog. 

\begin{figure*}[ht!]
\plotone{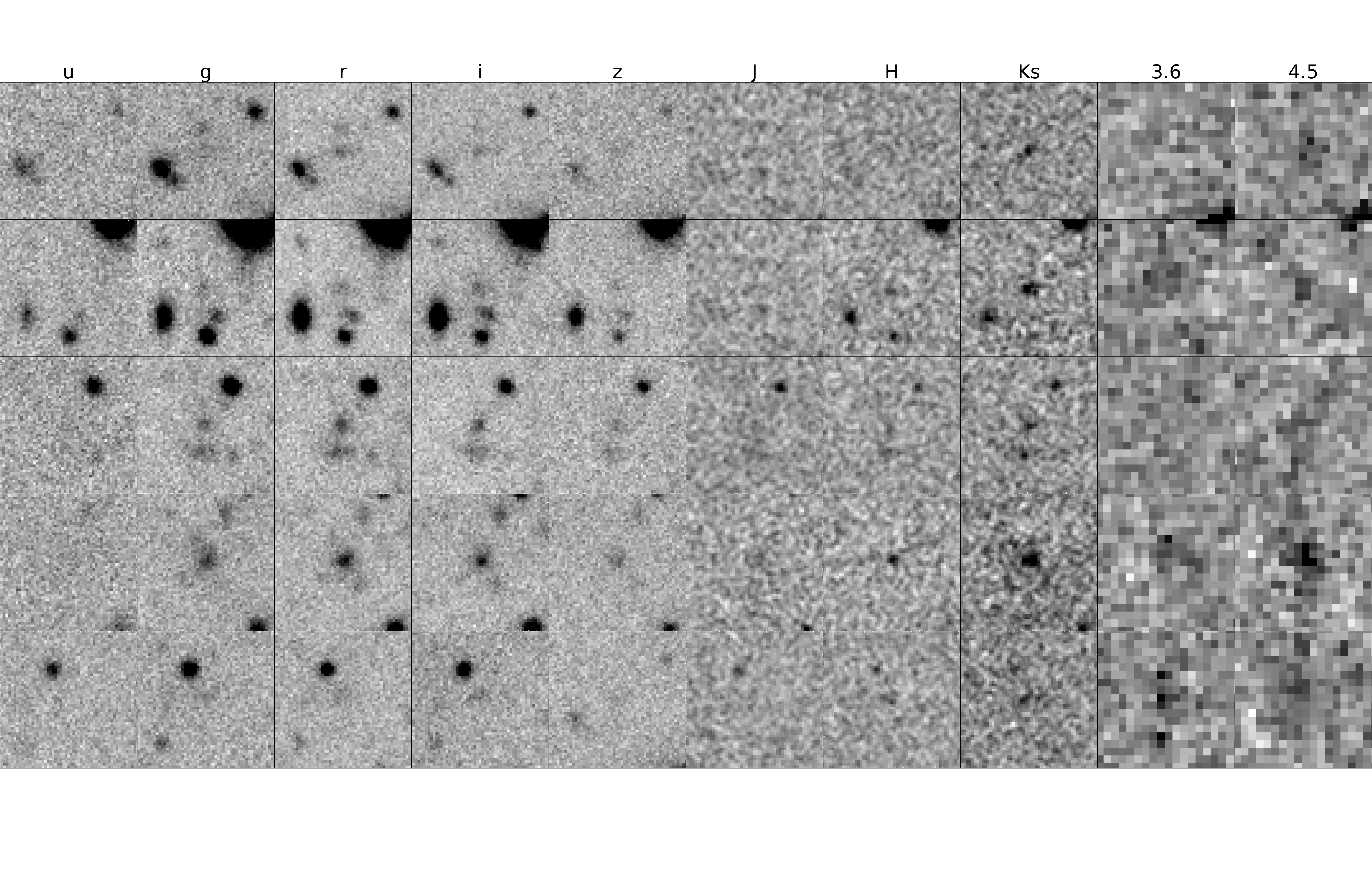}
\caption{
Example postage-stamp images of quiescent galaxies. All images are 10$\arcsec$ on each side. North is up and east is to the left.
}
\label{fig:images}
\end{figure*}

The right panel of Figure \ref{fig:D4000} shows the galaxies on the SFR--$\mathrm{M_{star}}$ plane where the galaxies are color-coded by their D4000 values. There is a clear trend that galaxies with larger D4000 tend to lie at the lower part of the plane. Indeed, all of our quiescent galaxies are located well below the star-forming main sequence relations at $z\sim3$ (only two are within 1 dex scatter of the relations while all the others are well beyond), further highlight the effectiveness and purity of using D4000 to select quiescent galaxies. Also seen from the figure is the lack of passive galaxies below $\sim10^{11}\mathrm{M_{\sun}}$, which is most likely due to selection effect. The quiescent galaxy candidates is 90\% complete above $10^{10.9}\mathrm{M_{\sun}}$, while the remaining star-forming galaxies has a completeness limit of $10^{10.7}\mathrm{M_{\sun}}$, as calculated using the method in Section \ref{sedfitting}.  This incompleteness issue also affects our results in Section \ref{dis}. A sample images of the quiescent galaxy candidates can be found in Figure \ref{fig:images}.

As a final check, in the left panel of Figure \ref{fig:cmass} we show our quiescent galaxy candidates in the $UVJ$ diagram. The rest-frame colors of the galaxies are derived from the best-fit templates from CIGALE. For secure determination of rest-frame $J$ band magnitude, we plot only the IRAC 3.6$\micron$ and 4.5$\micron$ detected sources (i.e., $>2\sigma$ magnitude limits).
Among the IRAC detected 29 candidates, 16 (55\%) are within the quiescent region defined by \cite{Muzzin13} while the remainder are also close to the quiescent parameter space. This further justifies our usage of D4000 index to select quiescent galaxies.

\begin{figure*}[ht!]
\plotone{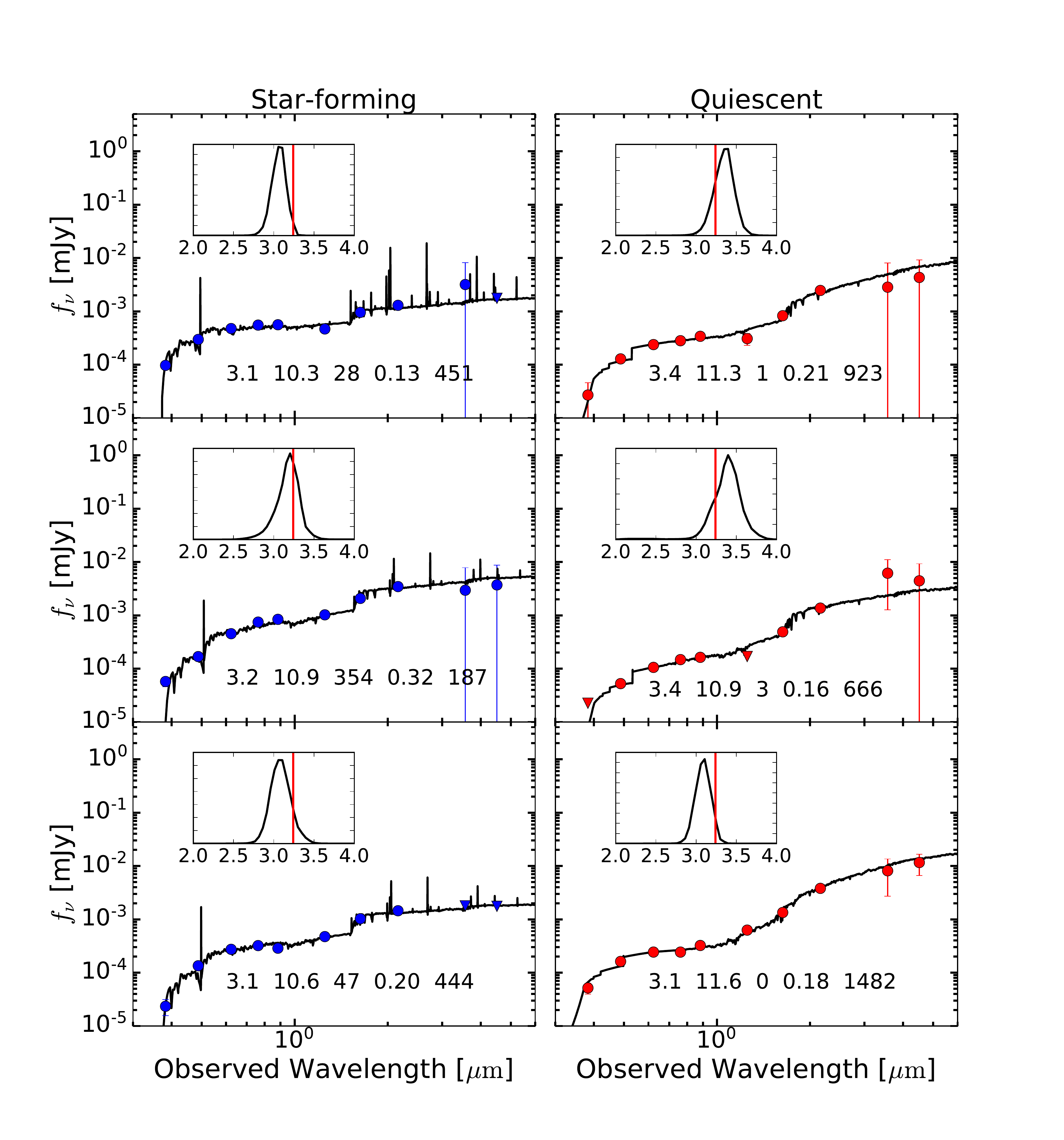}
\caption{
SED-fitting results for a subsample of the photo-z galaxies including star-forming galaxies and quiescent galaxies. The black solid lines are the best-fit model spectra. Filled circles represent the observed fluxes, while triangles denote 2$\sigma$ upper flux limits in the case of nondetection. In the inset of each panel, we also show the probability distribution function of the photometric redshift for each galaxy, and the
redshift of the protocluster is shown as a red vertical line. We also list the best-fit photo-$z$, log(M$_{star}$) (in units of M$_\sun$), SFR (in units of M$_\sun$ yr$^{-1}$), dust reddening parameter E(B–V), and age (in units of Myr) in the figure.
}
\label{fig:sed}
\end{figure*}

\begin{figure*}[ht!]
\plottwo{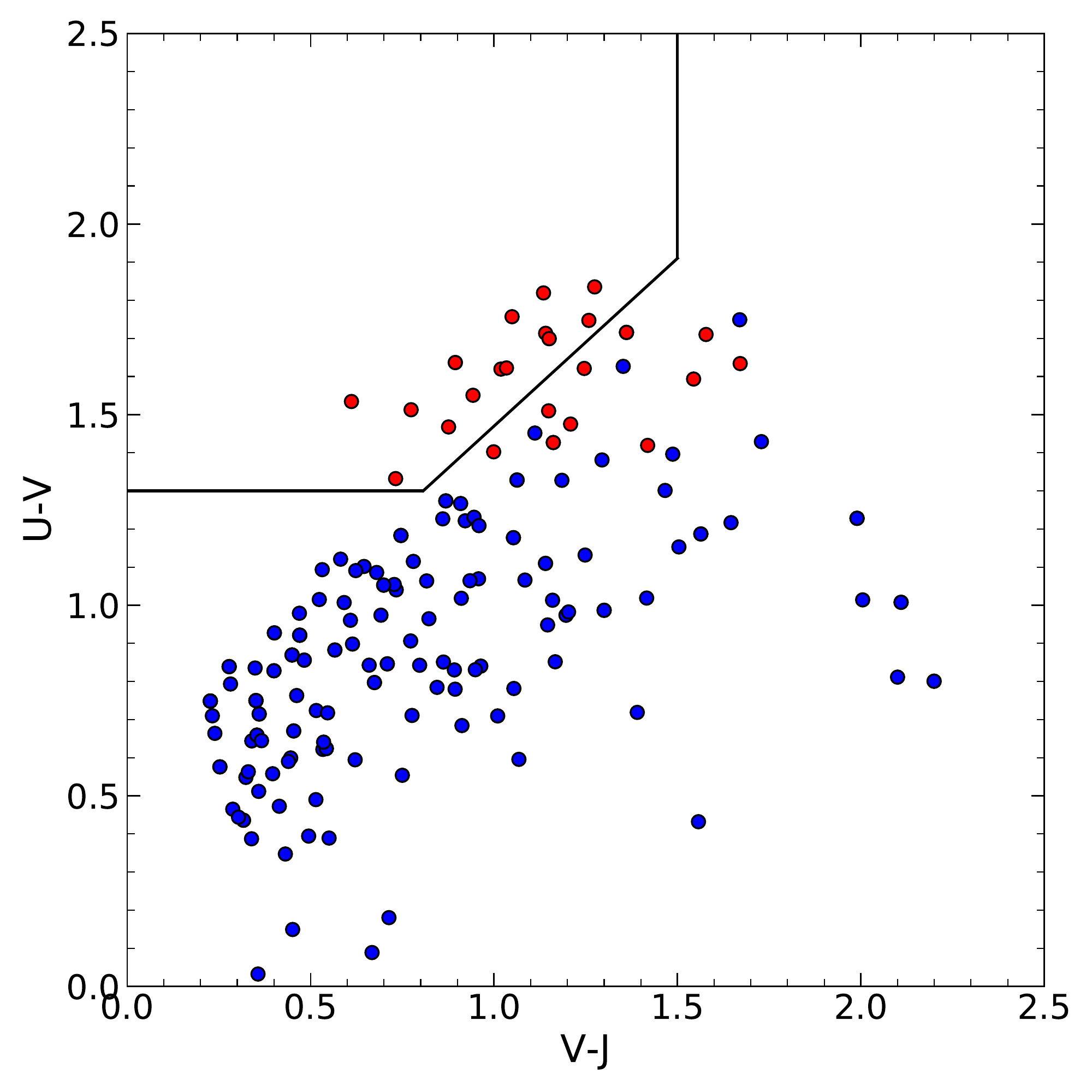}{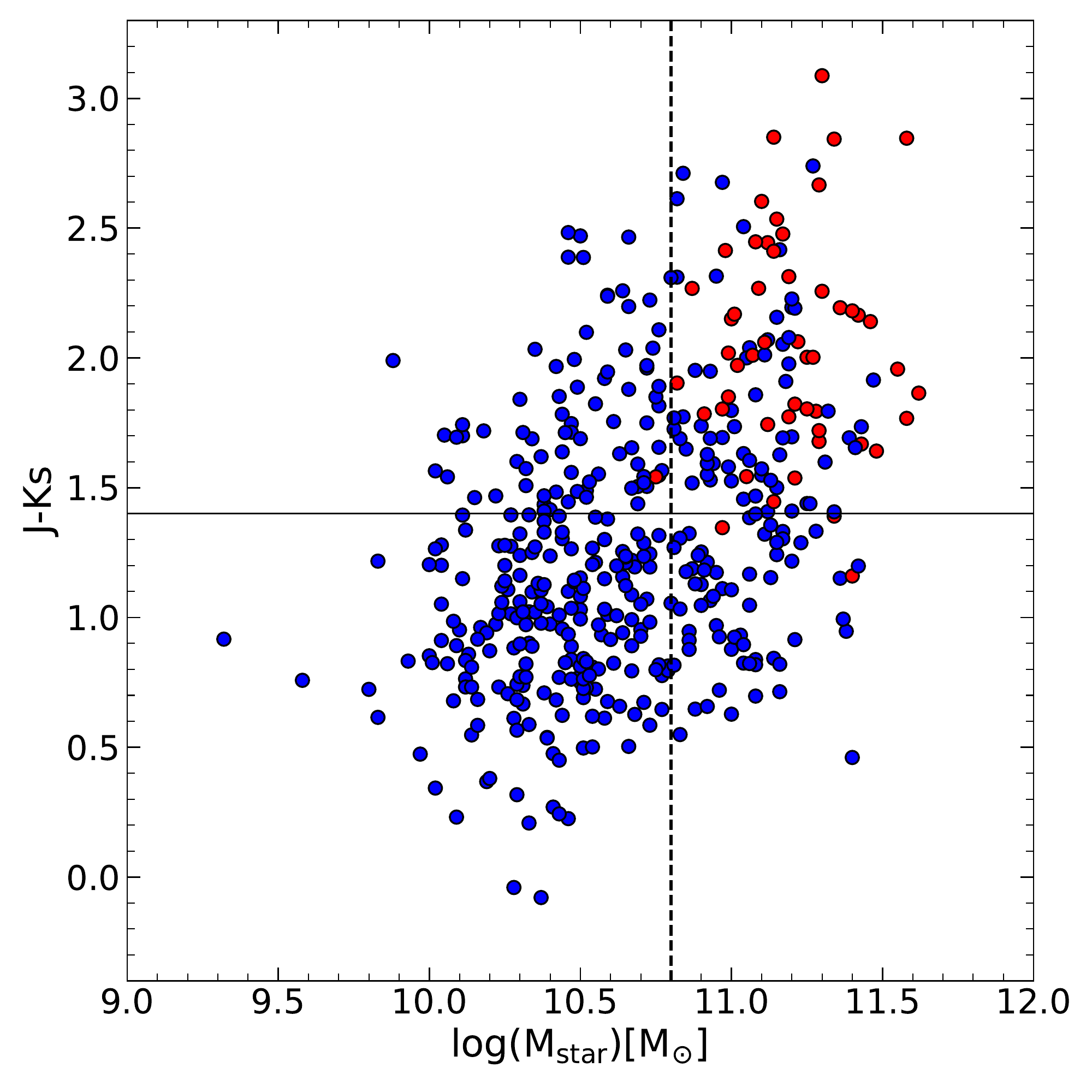}
\caption{
{\it Left:} Rest-frame $UVJ$ diagram for the IRAC detected sources. The red circles denote the quiescent galaxies defined by D4000 index while the blue circles are the star-forming galaxies. The solid lines are the cut used to define quiescent galaxies in \cite{Muzzin13}. Note that some galaxies have the same colors as their best-fit galaxy templates are the same (only at different redshifts).
{\it Right:} Mass-color diagram for all the photo-$z$ galaxies. The red circles denote the quiescent galaxies. The horizontal line represents the DRG criterion while the dashed vertical line is the stellar mass completeness limit of the sample.
}
\label{fig:cmass}
\end{figure*}

Figure \ref{fig:sed} shows the SED-fitting results for a subsample of the star-forming and quiescent galaxies defined using the above criterion. It can be seen that star-forming galaxies are featured by their prominant emission lines, and they are less massive ($<10^{11}\mathrm{M_\sun}$) and younger than the quiescent galaxies which are lack of nebular emissions. On the other hand, 80\% (42/52) of the quiescent galaxies have masses greater than $10^{11}\mathrm{M_\sun}$. Among the galaxies of masses $>10^{11}\mathrm{M_\sun}$, 38\% (42/112) are quiescent, which is similar to the quiescent fraction observed in \cite{Kubo13} and \cite{Ando20} in high-redshift protoclusters. 

It is noteworthy that these quiescent galaxies are very red with strong Balmer/4000\AA\ break between the $J$ and $K_S$ bands, with a median $J-K_S=2.0$.  In fact, 94\% (49/52) of the quiescent galaxies have $J-K_S>1.4$, which satisfy the Distant Red Galaxies (DRGs) selection criterion \citep{Franx03,van03}. DRGs are believed to be either dust obscured star-forming galaxies or old passive galaxies at $2<z<4$ \citep[e.g.,][]{Labbe05,Kriek06}. 
The right panel of Figure \ref{fig:cmass} shows the color-mass relation for the photo-$z$ galaxies. The quiescent galaxy candidates are concentrated in the top-right corner of the plane, suggesting they are among the most massive and reddest objects in the entire photo-$z$ sample. 

 However, we caution that whether all of these galaxies are truly ``red and dead'' remains uncertain. In lack of far-IR observations, especially the 24$\micron$ data, we are unable to quantify the possible emission features from polycyclic aromatic hydrocarbons \citep{Draine07} at rest-frame $\sim5.7\micron$ at $z=3.24$, which are heated by either dust obscured star-formation or AGNs. What is more, if some of these galaxies are dust-enshrouded star-forming galaxies, they could be detected at submillimeter wavelength by ALMA/SCUBA-2 \citep[e.g.,][]{Wang19}. At the current stage, without submillimeter observations it is difficult to further investigate this possibility. Nevertheless, we notice that \cite{Santini19} recently analysed in detail 26 candidate quiescent galaxies observed by ALMA at $3<z<5$ in the GOODS-South field. These galaxies were also selected using the SED-fitting technique from UV to mid-IR. They found none of these galaxies have secure detection ($>3\sigma$) in the submillimeter wavelength. Given the upper limits of the detection and with a stacking analysis, they found the dust obsecured star-formation activity is lower than that inferred from UV-optical. Meanwhile, using the ALMA-derived SFRs, $\sim$50\% of these galaxies are located at least 1$\sigma$ below the star-forming main sequence. They concluded that their sample is indeed quiescent in a statistical sense. Therefore, we argue that although we cannot completely rule out the contaminants of possible dusty star-forming galaxies, it is very unlikely that the red colors of all these candidates are caused by dust.

\section{sky distribution of galaxies}\label{distribution}
\subsection{Sky Distribution of the Photo-$z$ Galaxies} \label{photoz_dis}
Protoclusters are usually discovered as  overdensities of galaxies. In order to identify galaxy overdensities, many studies smoothed the spatial distributions of galaxies with a fixed or adaptive kernel to obtain the density maps \citep[e.g.,][]{Hayashino04,Matsuda05,Yang10,Lee14,Shi191,Shi192,Harikane19}.   Alternatively, some studies utilize a scale independent method named Voronoi tessellation \citep{Ramella01,Kim02,Cooper05,Soares11,Dey16} to measure the galaxy overdensity, and proved to be a good estimator of underlying density field \citep{Darvish15}. In this study, we use Voronoi tessellation to estimate the 2D surface density of the galaxies, which is described in the following.

A Voronoi tessellation is a unique way of dividing a two-dimensional distribution of points into convex cells, with each cell containing only one point and a set of vertices which are closer to that point than to any other in the plane. It has the property that the local density ($f$) of each cell is the inverse of the cell area ($a$). Therefore to estimate the overdensity of each cell, one first needs to calculate the average density of the cells in the entire plane ($\langle f \rangle=\langle 1/a \rangle$), then the density contrast of each cell is $\tilde{f}=f/\langle f \rangle$. 

The Voronoi tessellation of our photo-$z$ galaxies is shown in the top left panel of Figure \ref{fig:voronoi}. We can see that there is a large overdensity of galaxies in the western end of the field near the five spectroscopically confirmed LBGs at $z=3.24$ \citep{Toshikawa16}. As our sample is mass-limited and we may miss a lot of low-to-medium ($\sim10^{9}-10^{10}\mathrm{M_\sun}$) mass galaxies in the field that belong to the protocluster, therefore we define the protocluster region in a conservative way, with a circle enclosing all the spec-$z$ sources and most of the high-density ($\tilde{f}>1$) sources nearby. The average density of all the cells in the circle is $\tilde{f}=1.3$. The circle has a surface area of 81 arcmin$^2$ containing 96 galaxies, and its radius is $\sim10$ Mpc, consistent with the typical protocluster size at $z\sim3$ in \cite{Chiang13}. 

There are also several other overdensities in the field (the one in the southeast for example), but none of them is as significant as the large overdensity near the spec-$z$ sources within the circle. To verify this, we randomly put 100 circles with the same radius into the field and calculate the average density within. There are only four realizations in which the average density is comparable to the original one. The centers of these four circles are all very close to the one we used ($<4\arcmin$), and the average densities of the remaining realizations are all below 1.3. This confirms our visual impression and indicates the one near the spec-$z$ sources is indeed the largest overdensity in the field. The other small overdensities could be coincidental alignment of galaxies along the line of sight that have no physical
associations, or they could belong to smaller structures at $3.0<z<3.4$. At the current stage, without spectroscopic observations, we cannot determine which case is true, therefore we leave it to future studies. In the remainder of this paper, we only regard the one we defined in Figure \ref{fig:voronoi} as the protocluster region at $z=3.24$, and define the area outside the protocluster as the general field.

For comparison and completeness, in the bottom panel of Figure \ref{fig:voronoi} we show the smoothed surface density map of the $u$-dropout LBGs as in \cite{Toshikawa16}. We do not use Voronoi tessellation for the LBGs since there are nearly 6,000 sources which would make it difficult to identify overdense structures. We see there are two significant overdensities in the field.  The one in the north is roughly co-spatial with the five spectroscopic sources and the northeastern part of the photo-$z$ overdensity. The southern one is largely outside of the photo-$z$ overdensity and the southwestern part of the photo-$z$ overdensity disppears in the LBG map. The discrepancy between the photo-$z$ distribution and LBG distribution is not surprising, as they both have large redshift uncertainties ($\Delta z\sim0.2$ and $\Delta z\sim1$), which could dilute the genuine overdense structure and/or create fake density spikes along the line of sight. Alternatively, since they are selected in different ways (rest-frame optical vs. UV), and many of the UV selected LBGs are not detected in $K_S$ band (Section \ref{sedfitting}), photo-$z$ galaxies may represent more massive galaxy population that trace different underlying large-scale structures than the LBGs. Only future spectroscopic observations can verify these different scenarios.

The descendant mass of D4UD01 is calculated by \cite{Toshikawa16} using a set of lightcone models \citep{Henriques12} based on Millennium Simulation \citep{Springel05}.  They matched the observed surface density maps with those in the mock catalogs using the same selection method, finding a correlation between overdensity of LBGs and its descendant halo mass. The overdensity value of D4UD01 is 4.4, which results in a descendant halo mass of  $1.6\sim5\times10^{14}$~M$_\sun$ (see their Figure 7). According to \cite{Chiang13}, this structure will evolve into a Fornax-like (1--3$\times10^{14}$~M$_\sun$) or Virgo-like (3--10$\times10^{14}$~M$_\sun$) cluster at $z=0$. 

\subsection{Sky Distribution of the Quiescent Galaxies} \label{quiescent_dis}
Above we have shown that the survey field contains a large photo-$z$ galaxy overdensity, which is most likely to be a protocluster. In this section we investigate whether there is also presence of quiescent galaxies in this protocluster, which may shed light on how quenching of star formation depends on environment.

\begin{figure*}[ht!]
\plotone{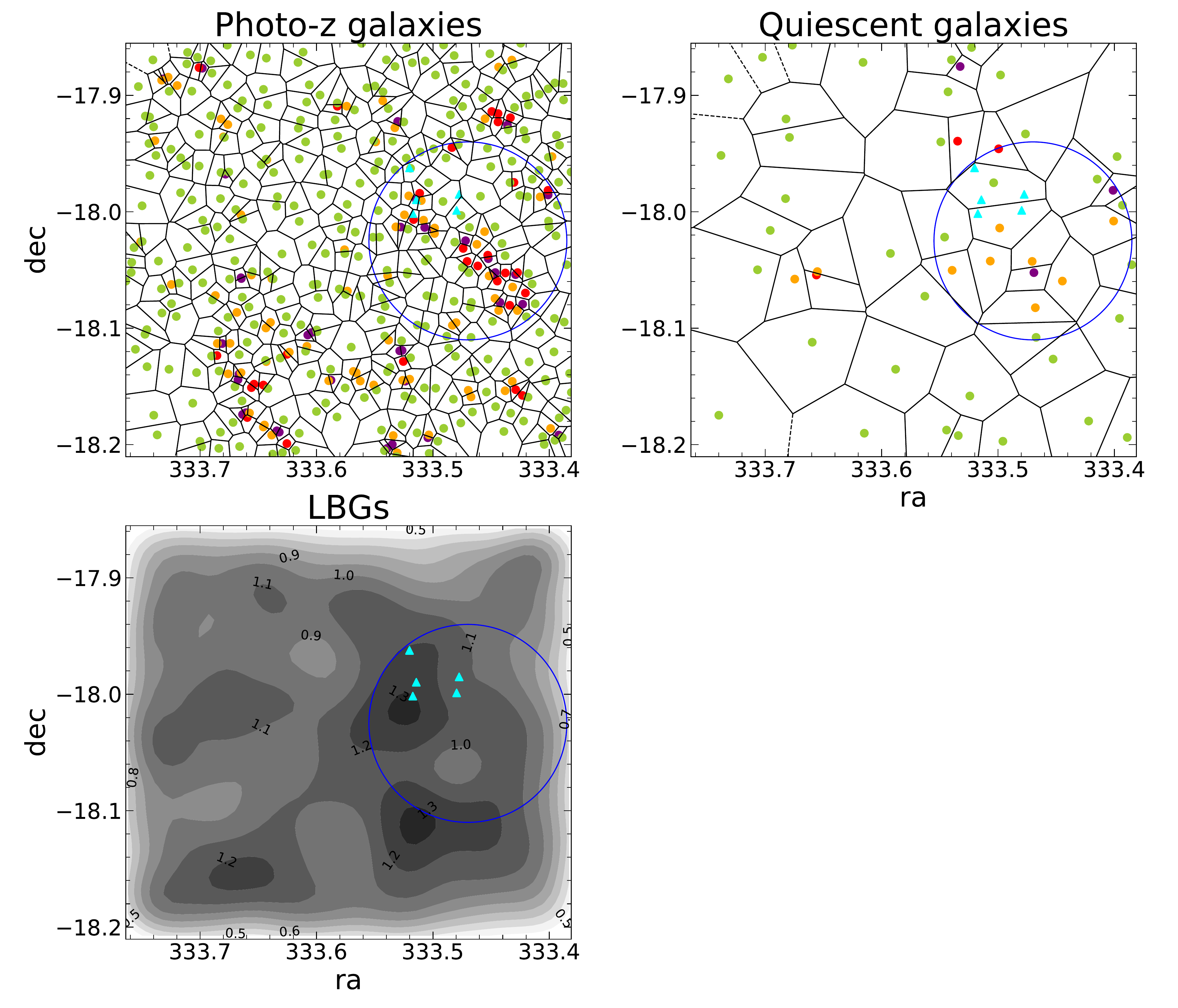}
\caption{{\it Top Left:} Voronoi tessellation of all the photo-$z$ galaxies in the field. The different colored circles respresent galaxies with different local density contrast (local density/mean density): $\tilde{f}>2.0$ (purple), $1.5<\tilde{f}<2.0$ (red), $1.0<\tilde{f}<1.5$ (orange), $\tilde{f}<1.0$ (green). The cyan triangles are the five spectroscopically confirmed LBGs at $z=3.24$. The large blue circle denotes the protocluster region. The dashed lines are associated with boarder cells that have infinite area, which are excluded from calculation of the average density.
{\it Top Right:} Voronoi tessellation of the quiescent galaxies. {\it Bottom left:} LBG density map smoothed using a FWHM=6 Mpc Gaussian kernel, and the contour labels show surface density levels relative to the field. }
\label{fig:voronoi}
\end{figure*}

The top right panel of Figure \ref{fig:voronoi} shows the Voronoi tessellation of the quiescent galaxies selected in Section \ref{sec3}. It is clear that the quiescent galaxies tend to be concentrated in the protocluster region: the galaxies within the circle have an average density of $\tilde{f}=1.4$ that is the largest in the entire field. There are 15 quiescent galaxies inside the protocluster region, resulting in a surface number density of $0.19$ arcmin$^{-2}$. In comparison, the surface density of all the quiescent galaxies in the entire field is 0.10 arcmin$^{-2}$ (52/497). Thus the number density of quiescent galaxy candidates in the protocluster nearly doubles that in the average field.
On the other hand, the surface density of all galaxies in the protocluster is 1.2 arcmin$^{-2}$, only mildly higher than the surface density of all galaxies in the entire field which is 0.9 arcmin$^{-2}$. Therefore the enhanced number of quiescent galaxies in the protocluster cannot be simply explained by the overall increased number of galaxies therein. In addition, the quiescent fraction in the protocluster is $\sim$16\%, which is also higher than that in the entire field of $\sim$12\%. These results clearly show that this protocluster contains a higher fraction of quiescent galaxies than the field, which cannot be explained by the larger number of galaxies in the overdense region. \footnote{Although we use D4000 to select quiescent galaxies in this work, this conclusion does not depend on specific selection criteria. For example, if instead we use sSFR$<10^{-11}$ yr$^{-1}$ as in \cite{Fontanot09} to define quiescent galaxies, we would have 53 quiescent galaxies among which 14 are inside the protocluster region. Therefore our results remain nearly the same.}

A large fraction of massive quiescent galaxies in this protocluster strongly suggests that cluster galaxies formed earlier than those in the field, that we may be witnessing the environmental quenching that takes place in the early stage of cluster formation long before virialization. These quiescent galaxies may have experienced an accelerated mass assembly in the high-density protocluster environment. We will discuss the environmental impacts on the galaxy stellar mass functions and star-formation activities in the next section, to further reveal the possible differences of galaxies' physical properties caused by the environments.

\section{Discussion} \label{dis}
\subsection{Environmental Dependence on the Galaxy Stellar Mass Function}
We have shown that this protocluster appears to host a higher fraction of massive quiescent galaxies than in the field. To further investigate the possible environmental trends in detail, we calculate the galaxy stellar mass function (GSMF) of the photo-$z$ galaxies in this section.

 As our survey volume is not large enough, cosmic variance (CV) might be a severe issue that affects the uncertainty of the number count. We use the Cosmic Variance Calculator \citep{Trenti08} to account for the CV, and the final error of each mass bin includes both Poission noise and the corresponding CV. To calculate the volume, we use the comoving volume at redshift $z=3.0-3.4$ with the corresponding survey area.
We do not account for the mass incompleteness in determining the GSMF, and it will not affect our major conclusion since our purpose is to compare the protocluster galaxies with the field galaxies in the same survey.

\begin{figure*}[ht!]
\plotone{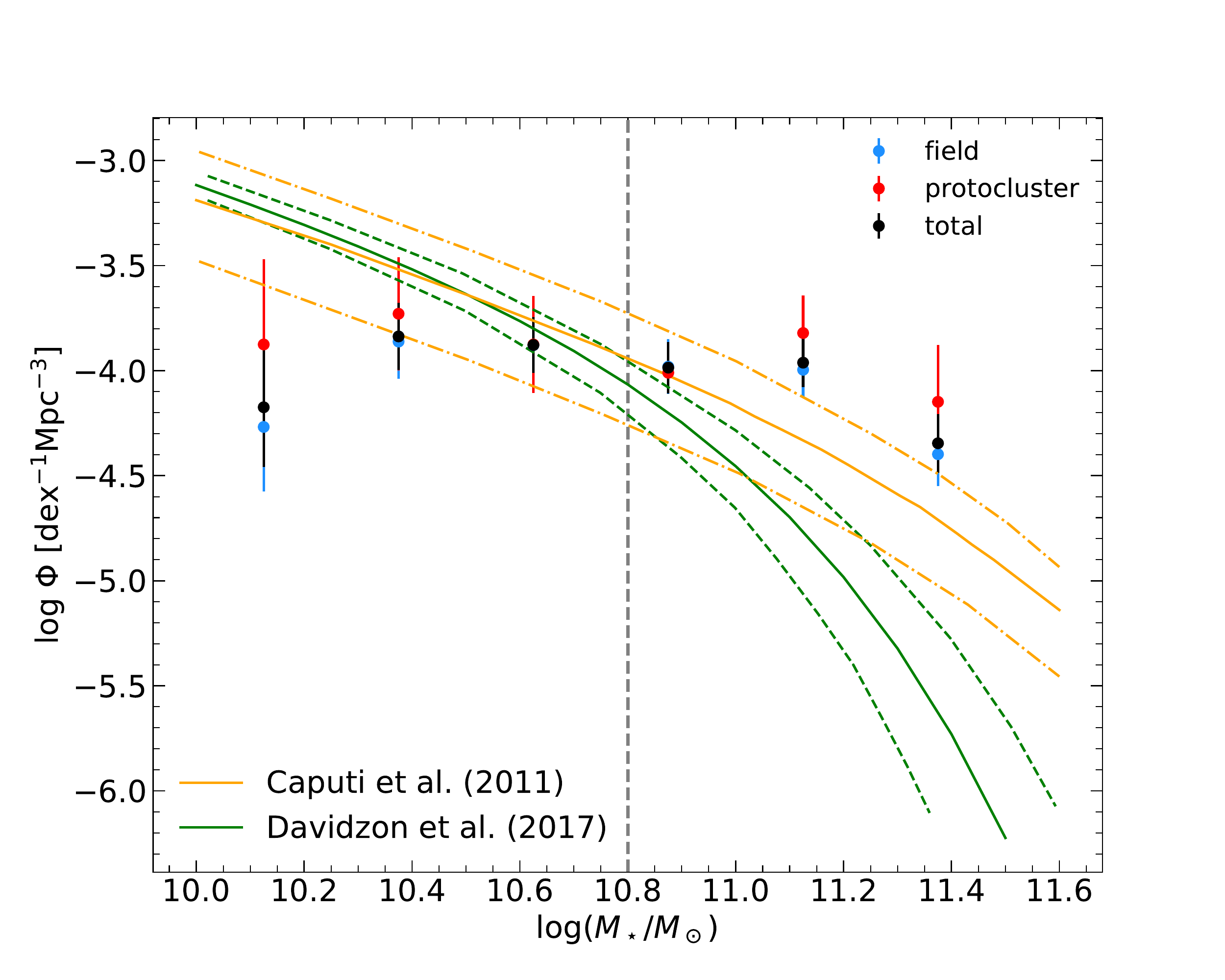}
\caption{
The GSMFs of different samples. The green and orange lines are the GSMFs from \cite{Davidzon17} and \cite{Caputi11} at $3.0<z<3.5$. The dashed and dash-dotted lines are the corresponding 1$\sigma$ uncertainties. The dashed vertical line represents the completeness limit of the sample.
}
\label{fig:massf}
\end{figure*}

Figure \ref{fig:massf} shows the GSMF of all the photo-$z$ galaxies including the protocluster and field galaxies. For comparison, we also plot the GSMFs from \cite{Caputi11} and \cite{Davidzon17} at the similar redshift range $3.0<z<3.5$ as our photo-$z$ galaxies. 

Above the mass completeness limit, our total GSMF agrees relatively well with \cite{Caputi11}. While the GSMF of  \cite{Davidzon17} has a faster decline at high-mass end, which was also noticed in \cite{Davidzon17} and was attributed to cosmic variance or difference in the photo-$z$ calculation. 
If we compare galaxies within our own sample, the selection effects could largely be ignored and our analysis would be more robust. From Figure \ref{fig:massf} we can see that protocluster galaxies appear to have increased number density than the field both at low-mass end ($\lesssim10^{10.6}\mathrm{M_{\sun}}$) and at high-mass end ($\gtrsim10^{11}\mathrm{M_{\sun}}$). While at stellar mass between $10^{10.6}\mathrm{M_{\sun}}$ and $10^{11}\mathrm{M_{\sun}}$ there is a sudden number drop of protocluster galaxies, making them almost indistinguishable from the field. This declined number density of protocluster galaxies at medium mass range  might be attributed to the incompleteness of the quiescent galaxy population. In Section \ref{quiescent_dis} we see that the protocluster region hosts a higher fraction of quiescent galaxies than the average field. Under this circumstance, we suspect that if quiescent galaxies are preferrably located in the protocluster, many would not be detected in our study below the mass completeness limit of $10^{10.9}\mathrm{M_{\sun}}$ (Section \ref{sec3}), resulting a `dip' in the medium mass range. If this is the case, it will imply an overall accelerated galaxy growth in the protocluster.

The above results suggest that we are witnessing an accelerated mass assembly in D4UD01. The fact that the protocluster hosts a higher fraction of quiescent galaxies also indicates a sign of environmental quenching. In addition, if the lack of medium mass galaxies in D4UD01 is due to the incompleteness of quiescent galaxies, the quiescent fraction will be even higher than that calculated in Section \ref{quiescent_dis}, making the quenching more effecient in D4UD01 than the field.

 Similar trends have also been found in many other protocluster studies. For example, \cite{Lemaux14,Lemaux18} discovered two protoclusters at $z=3.29$ and $z=4.57$ in the VIMOS Ultra-Deep Survey using spectroscopic observations. They found that these protoclusters tend to have an excess of more red and massive galaxies ralative to the coeval field.  Recently, \cite{Ando20} searched for protocluster cores using pairs of massive galxies at $z\sim2$ in the COSMOS field, finding that the core galaxies have a more top-heavy GSMF and a higher quiescent fraction than the field. In addition, \cite{Muldrew18} investigated galaxy evolution in protoclusters using a semi-analytic model from the Millennium Simulation and found the star formation histories of protocluster and field galaxies are very different. They argued this is because protoclusters have a high abundance of massive dark matter halos with top-heavy halo mass functions, which result in an early formation of massive galaxies and rapid merging of low-mass satellite galaxies followed by swift quenching. These independent studies reinforce the notion that protocluster galaxies experience  accelerated growth at high redshift and the cluster red sequence may already have been formed long before the final coalescence of the structure.

\subsection{Environmental Impacts on Physical Properties of Galaxies}
In this section, we compare the physical properties of galaxies in and out of the protocluster, to further discern possible environmental dependence on galaxy properties.

\begin{figure*}[ht!]
\plotone{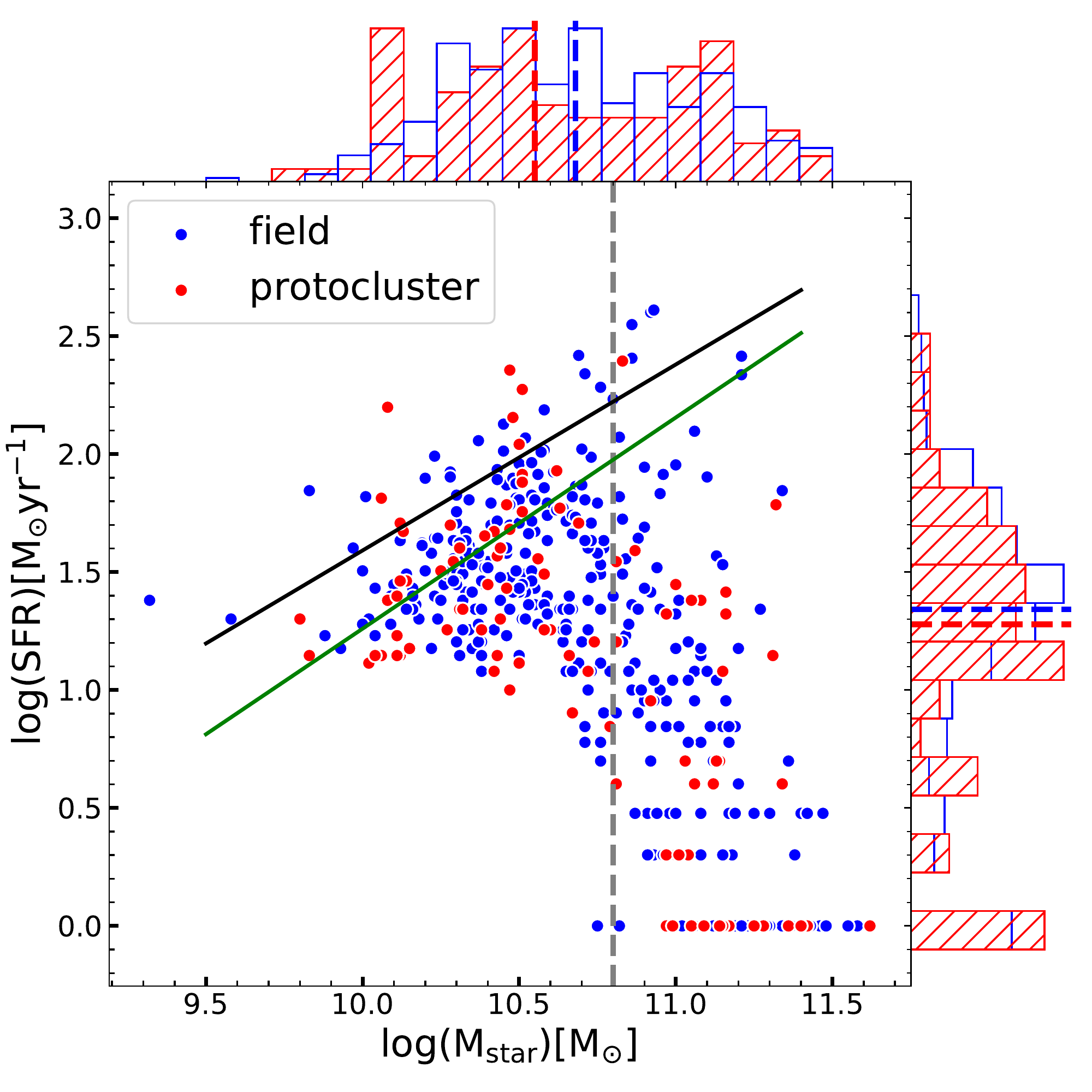}
\caption{
SFR--M$_\textrm{star}$ relation for the protocluster and field galaxies in our sample. The black solid line is the observed relation based on the calibration of \cite{Speagle14}, while the green solid line represents that from a semi-analytic model by \cite{Dutton10}. The vertical dashed line is the mass completeness limit of our sample. Galaxies with SFR=0 are indicated in the log(SFR)=0 location. The normalized histograms show the distributions of the SFR and stellar mass of the two groups, with the vertical lines indicating the median values.
}
\label{fig:ms}
\end{figure*}

 In Figure \ref{fig:ms} we show the photo-$z$ galaxies on the SFR-M$_\textrm{star}$ plane grouped by different environments. First, using the Kolmogorov-Smirnov (K–S) test, no significant differences ($p$-values$>$0.1) between the two groups in either the SFR or M$_\textrm{star}$ are found. It is possible that the large photo-$z$ uncertainty dilute the signal of potential differences in galaxy properties. The lack of medium mass galaxies in the protocluster is likely due to the combination of selection effect and environmental quenching as discussed earlier, which results in a lower median mass than the field galaxies. On the other hand, we notice that the SFRs of protocluster galaxies appear to be skewed towards lower values than the field counterparts, as can be seen in the histogram, suggesting possible suppresion of star-formation activites. This is also consistent with our previous findings of higher abundance of quiescent galaxies in this protocluster. Nevertheless, overall this trend is too weak to be recognized in the K-S test, and we tend to not give a definite conclusion here but leave it to future study when precise spectroscopic observations on this protocluster are available.
 
Many studies showed that the star-formation activities are enhanced in dense protocluster environments \citep[e.g.,][]{Koyama13,Hayashi16, Shimakawa18, Ito20, Shi20}. Although these findings shed light on the possible reversal of ``star formation--density'' relation in some protoclusters, there are many other protoclusters where no such differences are seen. For example, \cite{Cucciati14} studied a protocluster at $z=2.9$ in the COSMOS field and analyzed a spectroscopic sample of galaxies within the protocluster. When comparing with a control sample in the field, they could not identify any significant physical differences between the two samples. Similarly, no enhancement of star-formation have been found in two protoclusters at $z=3.29$ and $z=4.57$ in the VIMOS Ultra-Deep Survey \citep{Lemaux14,Lemaux18}. In addition, \cite{Shi191} analyzed a protocluster at $z=3.78$ using the similar photo-$z$ technique as in this work, and found no significant environmental impacts on star-formation activities. 

Although these different results sometimes appear to be contradictory, we argue this could be likely due to the different evolutionary stages and dynamical states these protoclusters are undergoing. For those protoclusters where enhancement of star-formation activities have been found, they could be experiencing an early mass assembly and  accelerated structure formation \citep{Steidel05}, resulting in the enhancement of star-formation we observed. While for those protoclusters where no such trends are spotted, they may have already passed the peak era of their star-formation or still in the early phase of formation before the emergence of any environmental effects \citep{Toshikawa14}. In this context, the differences observed in different protoclusters could originate from the so-called ``halo assembly bias'', in a sense that the properties of galaxies depend not only on the mass of the halo they reside in, but also on the halo formation time \citep[e.g.,][]{Gao05,Wechsler06,Li08,Zentner14}. It is noteworthy that \cite{Shi192} and \cite{Shi20} conducted a detailed study of a massive protocluster at $z=3.13$, finding the protocluster consists of two disjoint structures where one contains mostly low-mass star-forming galaxies while the other hosts a large fraction of massive quiescent and/or dusty galaxies. They also found that the former has a more enhanced star-formation activity than the latter while the latter is more similar to the field. All in all, these studies suggest that D4UD01 may be a more evolved structure that have already passed the peak of its star-formation era, so that we do not find any enhancement of star-formation but an excess of massive quiescent galaxies within.

So far, we have not considered how galaxies' dust content may change with the environment. Many studies have suggested that distant protoclusters often host extremely dusty star-forming galaxies such as submillimeter galaxies (SMGs) \citep[e.g.,][]{Casey16,Miller18,Umehata18,Cheng19}. These SMGs usually have extremely high star-formation rates ($>1000$ M$_\sun$/yr) and are generally invisible in rest-frame UV-NIR wavelengths due to heavily dust obscuration. If these dusty star-forming galaxies exist in our protocluster they would be totally missed in our selection. Future submillimeter observations of D4UD01 may give us further insight into how massive galaxies have formed in this protocluster.

\section{Summary and Conclusion}\label{sum}
In this paper, with the help of multiwavelength data, we study a protocluster D4UD01 at $z=3.24$ by identifying its member galaxies using SED fitting and photometric redshift. In the 497 arcmin$^2$ field which hosts the protocluster, 450 $K_S$ band detected candidate galaxies at $3.0<z_\textrm{phot}<3.4$ are selected as the photo-$z$ sample, reaching a mass completeness of $10^{10.8}$~M$_\sun$. We investigate their distributions in the field and probe possible environmental trends in the protocluster. Our main conclusions are summarized below.

1. Using D4000 index, 52 members are classified as quiescent galaxies in the photo-$z$ sample. Among these galaxies, 80\% have mass greater than $10^{11}$~M$_\sun$ and 94\% have colors consistent with those of DRGs. Therefore these galaxies are among the most massive and reddest ones in the entire sample.

2. A large galaxy overdensity is found in the field via Voronoi tessellation, which contains 96 sources. Being the largest overdensity in the entire field, we define this overdensity as the protocluster region. Interestingly, we find that the quiescent galaxies are mostly concentrated in the protocluster region with a higher quiescent fraction, suggesting potential environmental quenching effect is taking place in this protocluster.

3. The mass function of protocluster galaxies shows an enhancement in comparison to the field, suggesting an accelerated mass assembly in the protocluster. When further studying the environmental impacts on galaxy properties, a weak signal of suppressed star-formation activities is found in the protocluster comparing with the field, but the differences are not significant enough to be conclusive. It is argued that D4UD01 is a more evolved structure that already passed its peak star-formation era, than those younger protoclusters where enhanced star-formation activities were found.

We thank the anonymous referee for a careful review of this paper. K.S. is grateful to Richard Bielby and Patrick Petitjean for sharing the AAOmega spectroscopic data in the D4 field for photo-$z$ calibration. T.F. and K.S. acknowledge the funding from the National Key R\&D Program of China No. 2017YFA0402600, and NSFC grants No. 11525312, 11890692. K.S. is supported by NSFC grants No. 12003023 and China Postdoctoral Science Foundation (2020M680086). The CFHTLS data used in this work are based on observations obtained with MegaPrime/MegaCam, a joint project of CFHT and CEA/IRFU, at the Canada-France-Hawaii Telescope (CFHT) which is operated by the National Research Council (NRC) of Canada, the Institut National des Science de l'Univers of the Centre National de la Recherche Scientifique (CNRS) of France, and the University of Hawaii. This work is based in part on data products produced at Terapix available at the Canadian Astronomy Data Centre as part of the Canada-France-Hawaii Telescope Legacy Survey, a collaborative project of NRC and CNRS. The IRAC data used in this study are based on data and catalog products from NMBS-II IRAC, funded by the National Aeronautics and Space Administration (NASA) under grant number NNX16AN49G issued through the NNH15ZDA001N Astrophysics Data Analysis Program (ADAP).

\bibliography{xmu}

\begin{thebibliography}{}
\expandafter\ifx\csname natexlab\endcsname\relax\def\natexlab#1{#1}\fi
\providecommand{\url}[1]{\href{#1}{#1}}
\providecommand{\dodoi}[1]{doi:~\href{http://doi.org/#1}{\nolinkurl{#1}}}
\providecommand{\doeprint}[1]{\href{http://ascl.net/#1}{\nolinkurl{http://ascl.net/#1}}}
\providecommand{\doarXiv}[1]{\href{https://arxiv.org/abs/#1}{\nolinkurl{https://arxiv.org/abs/#1}}}

\bibitem[{{Alberts} {et~al.}(2014){Alberts}, {Pope}, {Brodwin}, {Atlee}, {Lin},
  {Dey}, {Eisenhardt}, {Gettings}, {Gonzalez}, {Jannuzi}, {Mancone},
  {Moustakas}, {Snyder}, {Stanford}, {Stern}, {Weiner}, \&
  {Zeimann}}]{Alberts14}
{Alberts}, S., {Pope}, A., {Brodwin}, M., {et~al.} 2014, \mnras, 437, 437,
  \dodoi{10.1093/mnras/stt1897}

\bibitem[{{Ando} {et~al.}(2020){Ando}, {Shimasaku}, \& {Momose}}]{Ando20}
{Ando}, M., {Shimasaku}, K., \& {Momose}, R. 2020, \mnras, 496, 3169,
  \dodoi{10.1093/mnras/staa1757}

\bibitem[{{Annunziatella} {et~al.}(2018){Annunziatella}, {Marchesini},
  {Stefanon}, {Muzzin}, {Lange-Vagle}, {Cybulski}, {Labbe}, {Kado-Fong},
  {Bezanson}, {Brammer}, {Herrera}, {Lundgren}, {Marsan}, {Nonino}, {Rudnick},
  {Saracco}, {Tomer}, {Valdes}, {van der Burg}, {van Dokkum}, {Wake}, \&
  {Whitaker}}]{Annunziatella18}
{Annunziatella}, M., {Marchesini}, D., {Stefanon}, M., {et~al.} 2018, \pasp,
  130, 124501, \dodoi{10.1088/1538-3873/aae796}

\bibitem[{{Balogh} {et~al.}(1999){Balogh}, {Morris}, {Yee}, {Carlberg}, \&
  {Ellingson}}]{Balogh99}
{Balogh}, M.~L., {Morris}, S.~L., {Yee}, H.~K.~C., {Carlberg}, R.~G., \&
  {Ellingson}, E. 1999, \apj, 527, 54, \dodoi{10.1086/308056}

\bibitem[{{Bertin} \& {Arnouts}(1996)}]{Bertin96}
{Bertin}, E., \& {Arnouts}, S. 1996, \aaps, 117, 393,
  \dodoi{10.1051/aas:1996164}

\bibitem[{{Bielby} {et~al.}(2012){Bielby}, {Hudelot}, {McCracken}, {Ilbert},
  {Daddi}, {Le F{\`e}vre}, {Gonzalez-Perez}, {Kneib}, {Marmo}, {Mellier},
  {Salvato}, {Sanders}, \& {Willott}}]{Bielby12}
{Bielby}, R., {Hudelot}, P., {McCracken}, H.~J., {et~al.} 2012, \aap, 545, A23,
  \dodoi{10.1051/0004-6361/201118547}

\bibitem[{{Boquien} {et~al.}(2019){Boquien}, {Burgarella}, {Roehlly}, {Buat},
  {Ciesla}, {Corre}, {Inoue}, \& {Salas}}]{Boquien19}
{Boquien}, M., {Burgarella}, D., {Roehlly}, Y., {et~al.} 2019, \aap, 622, A103,
  \dodoi{10.1051/0004-6361/201834156}

\bibitem[{{Bower} {et~al.}(1992){Bower}, {Lucey}, \& {Ellis}}]{Bower92}
{Bower}, R.~G., {Lucey}, J.~R., \& {Ellis}, R.~S. 1992, \mnras, 254, 601,
  \dodoi{10.1093/mnras/254.4.601}

\bibitem[{{Brammer} {et~al.}(2011){Brammer}, {Whitaker}, {van Dokkum},
  {Marchesini}, {Franx}, {Kriek}, {Labb{\'e}}, {Lee}, {Muzzin}, {Quadri},
  {Rudnick}, \& {Williams}}]{Brammer11}
{Brammer}, G.~B., {Whitaker}, K.~E., {van Dokkum}, P.~G., {et~al.} 2011, \apj,
  739, 24, \dodoi{10.1088/0004-637X/739/1/24}

\bibitem[{{Brinchmann} {et~al.}(2004){Brinchmann}, {Charlot}, {White},
  {Tremonti}, {Kauffmann}, {Heckman}, \& {Brinkmann}}]{Brinchmann04}
{Brinchmann}, J., {Charlot}, S., {White}, S.~D.~M., {et~al.} 2004, \mnras, 351,
  1151, \dodoi{10.1111/j.1365-2966.2004.07881.x}

\bibitem[{{Bruzual} \& {Charlot}(2003)}]{BC03}
{Bruzual}, G., \& {Charlot}, S. 2003, \mnras, 344, 1000,
  \dodoi{10.1046/j.1365-8711.2003.06897.x}

\bibitem[{{Cai} {et~al.}(2017){Cai}, {Fan}, {Yang}, {Bian}, {Prochaska},
  {Zabludoff}, {McGreer}, {Zheng}, {Green}, {Cantalupo}, {Frye}, {Hamden},
  {Jiang}, {Kashikawa}, \& {Wang}}]{Cai17}
{Cai}, Z., {Fan}, X., {Yang}, Y., {et~al.} 2017, \apj, 837, 71,
  \dodoi{10.3847/1538-4357/aa5d14}

\bibitem[{{Calzetti} {et~al.}(2000){Calzetti}, {Armus}, {Bohlin}, {Kinney},
  {Koornneef}, \& {Storchi-Bergmann}}]{Calzetti00}
{Calzetti}, D., {Armus}, L., {Bohlin}, R.~C., {et~al.} 2000, \apj, 533, 682,
  \dodoi{10.1086/308692}

\bibitem[{{Caputi} {et~al.}(2011){Caputi}, {Cirasuolo}, {Dunlop}, {McLure},
  {Farrah}, \& {Almaini}}]{Caputi11}
{Caputi}, K.~I., {Cirasuolo}, M., {Dunlop}, J.~S., {et~al.} 2011, \mnras, 413,
  162, \dodoi{10.1111/j.1365-2966.2010.18118.x}

\bibitem[{{Casey}(2016)}]{Casey16}
{Casey}, C.~M. 2016, \apj, 824, 36, \dodoi{10.3847/0004-637X/824/1/36}

\bibitem[{{Chabrier}(2003)}]{Chabrier03}
{Chabrier}, G. 2003, \pasp, 115, 763, \dodoi{10.1086/376392}

\bibitem[{{Cheng} {et~al.}(2019){Cheng}, {Clements}, {Greenslade}, {Cairns},
  {Andreani}, {Bremer}, {Conversi}, {Cooray}, {Dannerbauer}, {De Zotti},
  {Eales}, {Gonz{\'a}lez-Nuevo}, {Ibar}, {Leeuw}, {Ma}, {Micha{\l}owski},
  {Nayyeri}, {Riechers}, {Scott}, {Temi}, {Vaccari}, {Valtchanov}, {van
  Kampen}, \& {Wang}}]{Cheng19}
{Cheng}, T., {Clements}, D.~L., {Greenslade}, J., {et~al.} 2019, \mnras, 490,
  3840, \dodoi{10.1093/mnras/stz2640}

\bibitem[{{Chiang} {et~al.}(2013){Chiang}, {Overzier}, \&
  {Gebhardt}}]{Chiang13}
{Chiang}, Y.-K., {Overzier}, R., \& {Gebhardt}, K. 2013, \apj, 779, 127,
  \dodoi{10.1088/0004-637X/779/2/127}

\bibitem[{{Chiang} {et~al.}(2017){Chiang}, {Overzier}, {Gebhardt}, \&
  {Henriques}}]{Chiang17}
{Chiang}, Y.-K., {Overzier}, R.~A., {Gebhardt}, K., \& {Henriques}, B. 2017,
  \apjl, 844, L23, \dodoi{10.3847/2041-8213/aa7e7b}

\bibitem[{{Cooper} {et~al.}(2005){Cooper}, {Newman}, {Madgwick}, {Gerke},
  {Yan}, \& {Davis}}]{Cooper05}
{Cooper}, M.~C., {Newman}, J.~A., {Madgwick}, D.~S., {et~al.} 2005, \apj, 634,
  833, \dodoi{10.1086/432868}

\bibitem[{{Cooper} {et~al.}(2008){Cooper}, {Newman}, {Weiner}, {Yan},
  {Willmer}, {Bundy}, {Coil}, {Conselice}, {Davis}, {Faber}, {Gerke},
  {Guhathakurta}, {Koo}, \& {Noeske}}]{Cooper08}
{Cooper}, M.~C., {Newman}, J.~A., {Weiner}, B.~J., {et~al.} 2008, \mnras, 383,
  1058, \dodoi{10.1111/j.1365-2966.2007.12613.x}

\bibitem[{{Cucciati} {et~al.}(2014){Cucciati}, {Zamorani}, {Lemaux},
  {Bardelli}, {Cimatti}, {Le F{\`e}vre}, {Cassata}, {Garilli}, {Le Brun},
  {Maccagni}, {Pentericci}, {Tasca}, {Thomas}, {Vanzella}, {Zucca}, {Amorin},
  {Capak}, {Cassar{\`a}}, {Castellano}, {Cuby}, {de la Torre}, {Durkalec},
  {Fontana}, {Giavalisco}, {Grazian}, {Hathi}, {Ilbert}, {Moreau}, {Paltani},
  {Ribeiro}, {Salvato}, {Schaerer}, {Scodeggio}, {Sommariva}, {Talia},
  {Taniguchi}, {Tresse}, {Vergani}, {Wang}, {Charlot}, {Contini}, {Fotopoulou},
  {L{\'o}pez-Sanjuan}, {Mellier}, \& {Scoville}}]{Cucciati14}
{Cucciati}, O., {Zamorani}, G., {Lemaux}, B.~C., {et~al.} 2014, \aap, 570, A16,
  \dodoi{10.1051/0004-6361/201423811}

\bibitem[{{Dale} {et~al.}(2014){Dale}, {Helou}, {Magdis}, {Armus},
  {D{\'\i}az-Santos}, \& {Shi}}]{Dale14}
{Dale}, D.~A., {Helou}, G., {Magdis}, G.~E., {et~al.} 2014, \apj, 784, 83,
  \dodoi{10.1088/0004-637X/784/1/83}

\bibitem[{{Darvish} {et~al.}(2015){Darvish}, {Mobasher}, {Sobral}, {Scoville},
  \& {Aragon-Calvo}}]{Darvish15}
{Darvish}, B., {Mobasher}, B., {Sobral}, D., {Scoville}, N., \& {Aragon-Calvo},
  M. 2015, \apj, 805, 121, \dodoi{10.1088/0004-637X/805/2/121}

\bibitem[{{Davidzon} {et~al.}(2017){Davidzon}, {Ilbert}, {Laigle}, {Coupon},
  {McCracken}, {Delvecchio}, {Masters}, {Capak}, {Hsieh}, {Le F{\`e}vre},
  {Tresse}, {Bethermin}, {Chang}, {Faisst}, {Le Floc'h}, {Steinhardt}, {Toft},
  {Aussel}, {Dubois}, {Hasinger}, {Salvato}, {Sanders}, {Scoville}, \&
  {Silverman}}]{Davidzon17}
{Davidzon}, I., {Ilbert}, O., {Laigle}, C., {et~al.} 2017, \aap, 605, A70,
  \dodoi{10.1051/0004-6361/201730419}

\bibitem[{{Dey} {et~al.}(2016){Dey}, {Lee}, {Reddy}, {Cooper}, {Inami}, {Hong},
  {Gonzalez}, \& {Jannuzi}}]{Dey16}
{Dey}, A., {Lee}, K.-S., {Reddy}, N., {et~al.} 2016, \apj, 823, 11,
  \dodoi{10.3847/0004-637X/823/1/11}

\bibitem[{{Draine} \& {Li}(2007)}]{Draine07}
{Draine}, B.~T., \& {Li}, A. 2007, \apj, 657, 810, \dodoi{10.1086/511055}

\bibitem[{{Dressler}(1980)}]{Dressler80}
{Dressler}, A. 1980, \apj, 236, 351, \dodoi{10.1086/157753}

\bibitem[{{Dressler} {et~al.}(1997){Dressler}, {Oemler}, {Couch}, {Smail},
  {Ellis}, {Barger}, {Butcher}, {Poggianti}, \& {Sharples}}]{Dressler97}
{Dressler}, A., {Oemler}, Augustus, J., {Couch}, W.~J., {et~al.} 1997, \apj,
  490, 577, \dodoi{10.1086/304890}

\bibitem[{{Dutton} {et~al.}(2010){Dutton}, {van den Bosch}, \&
  {Dekel}}]{Dutton10}
{Dutton}, A.~A., {van den Bosch}, F.~C., \& {Dekel}, A. 2010, \mnras, 405,
  1690, \dodoi{10.1111/j.1365-2966.2010.16620.x}

\bibitem[{{Elbaz} {et~al.}(2007){Elbaz}, {Daddi}, {Le Borgne}, {Dickinson},
  {Alexander}, {Chary}, {Starck}, {Brand t}, {Kitzbichler}, {MacDonald},
  {Nonino}, {Popesso}, {Stern}, \& {Vanzella}}]{Elbaz07}
{Elbaz}, D., {Daddi}, E., {Le Borgne}, D., {et~al.} 2007, \aap, 468, 33,
  \dodoi{10.1051/0004-6361:20077525}

\bibitem[{{Fontanot} {et~al.}(2009){Fontanot}, {De Lucia}, {Monaco},
  {Somerville}, \& {Santini}}]{Fontanot09}
{Fontanot}, F., {De Lucia}, G., {Monaco}, P., {Somerville}, R.~S., \&
  {Santini}, P. 2009, \mnras, 397, 1776,
  \dodoi{10.1111/j.1365-2966.2009.15058.x}

\bibitem[{{Franx} {et~al.}(2003){Franx}, {Labb{\'e}}, {Rudnick}, {van Dokkum},
  {Daddi}, {F{\"o}rster Schreiber}, {Moorwood}, {Rix}, {R{\"o}ttgering}, {van
  der Wel}, {van der Werf}, \& {van Starkenburg}}]{Franx03}
{Franx}, M., {Labb{\'e}}, I., {Rudnick}, G., {et~al.} 2003, \apjl, 587, L79,
  \dodoi{10.1086/375155}

\bibitem[{{Gallazzi} {et~al.}(2005){Gallazzi}, {Charlot}, {Brinchmann},
  {White}, \& {Tremonti}}]{Gallazzi05}
{Gallazzi}, A., {Charlot}, S., {Brinchmann}, J., {White}, S. D.~M., \&
  {Tremonti}, C.~A. 2005, \mnras, 362, 41,
  \dodoi{10.1111/j.1365-2966.2005.09321.x}

\bibitem[{{Gao} {et~al.}(2005){Gao}, {Springel}, \& {White}}]{Gao05}
{Gao}, L., {Springel}, V., \& {White}, S. D.~M. 2005, \mnras, 363, L66,
  \dodoi{10.1111/j.1745-3933.2005.00084.x}

\bibitem[{{Giavalisco}(2002)}]{Giavalisco02}
{Giavalisco}, M. 2002, \araa, 40, 579,
  \dodoi{10.1146/annurev.astro.40.121301.111837}

\bibitem[{{Girelli} {et~al.}(2019){Girelli}, {Bolzonella}, \&
  {Cimatti}}]{Girelli19}
{Girelli}, G., {Bolzonella}, M., \& {Cimatti}, A. 2019, arXiv e-prints,
  arXiv:1910.07544.
\newblock \doarXiv{1910.07544}

\bibitem[{{Goto} {et~al.}(2003){Goto}, {Yamauchi}, {Fujita}, {Okamura},
  {Sekiguchi}, {Smail}, {Bernardi}, \& {Gomez}}]{Goto03}
{Goto}, T., {Yamauchi}, C., {Fujita}, Y., {et~al.} 2003, \mnras, 346, 601,
  \dodoi{10.1046/j.1365-2966.2003.07114.x}

\bibitem[{{Gwyn}(2012)}]{Gwyn12}
{Gwyn}, S. D.~J. 2012, \aj, 143, 38, \dodoi{10.1088/0004-6256/143/2/38}

\bibitem[{{Haines} {et~al.}(2017){Haines}, {Iovino}, {Krywult}, {Guzzo},
  {Davidzon}, {Bolzonella}, {Garilli}, {Scodeggio}, {Granett}, {de la Torre},
  {De Lucia}, {Abbas}, {Adami}, {Arnouts}, {Bottini}, {Cappi}, {Cucciati},
  {Franzetti}, {Fritz}, {Gargiulo}, {Le Brun}, {Le F{\`e}vre}, {Maccagni},
  {Ma{\l}ek}, {Marulli}, {Moutard}, {Polletta}, {Pollo}, {Tasca}, {Tojeiro},
  {Vergani}, {Zanichelli}, {Zamorani}, {Bel}, {Branchini}, {Coupon}, {Ilbert},
  {Moscardini}, {Peacock}, \& {Siudek}}]{Haines17}
{Haines}, C.~P., {Iovino}, A., {Krywult}, J., {et~al.} 2017, \aap, 605, A4,
  \dodoi{10.1051/0004-6361/201630118}

\bibitem[{{Harikane} {et~al.}(2019){Harikane}, {Ouchi}, {Ono}, {Fujimoto},
  {Donevski}, {Shibuya}, {Faisst}, {Goto}, {Hatsukade}, {Kashikawa}, {Kohno},
  {Hashimoto}, {Higuchi}, {Inoue}, {Lin}, {Martin}, {Overzier}, {Smail},
  {Toshikawa}, {Umehata}, {Ao}, {Chapman}, {Clements}, {Im}, {Jing},
  {Kawaguchi}, {Lee}, {Lee}, {Lin}, {Matsuoka}, {Marinello}, {Nagao},
  {Onodera}, {Toft}, \& {Wang}}]{Harikane19}
{Harikane}, Y., {Ouchi}, M., {Ono}, Y., {et~al.} 2019, \apj, 883, 142,
  \dodoi{10.3847/1538-4357/ab2cd5}

\bibitem[{{Hathi} {et~al.}(2009){Hathi}, {Ferreras}, {Pasquali}, {Malhotra},
  {Rhoads}, {Pirzkal}, {Windhorst}, \& {Xu}}]{Hathi09}
{Hathi}, N.~P., {Ferreras}, I., {Pasquali}, A., {et~al.} 2009, \apj, 690, 1866,
  \dodoi{10.1088/0004-637X/690/2/1866}

\bibitem[{{Hayashi} {et~al.}(2016){Hayashi}, {Kodama}, {Tanaka}, {Shimakawa},
  {Koyama}, {Tadaki}, {Suzuki}, \& {Yamamoto}}]{Hayashi16}
{Hayashi}, M., {Kodama}, T., {Tanaka}, I., {et~al.} 2016, \apjl, 826, L28,
  \dodoi{10.3847/2041-8205/826/2/L28}

\bibitem[{{Hayashino} {et~al.}(2004){Hayashino}, {Matsuda}, {Tamura},
  {Yamauchi}, {Yamada}, {Ajiki}, {Fujita}, {Murayama}, {Nagao}, {Ohta},
  {Okamura}, {Ouchi}, {Shimasaku}, {Shioya}, \& {Taniguchi}}]{Hayashino04}
{Hayashino}, T., {Matsuda}, Y., {Tamura}, H., {et~al.} 2004, \aj, 128, 2073,
  \dodoi{10.1086/424935}

\bibitem[{{Henriques} {et~al.}(2012){Henriques}, {White}, {Lemson}, {Thomas},
  {Guo}, {Marleau}, \& {Overzier}}]{Henriques12}
{Henriques}, B. M.~B., {White}, S. D.~M., {Lemson}, G., {et~al.} 2012, \mnras,
  421, 2904, \dodoi{10.1111/j.1365-2966.2012.20521.x}

\bibitem[{{Hinshaw} {et~al.}(2013){Hinshaw}, {Larson}, {Komatsu}, {Spergel},
  {Bennett}, {Dunkley}, {Nolta}, {Halpern}, {Hill}, {Odegard}, {Page}, {Smith},
  {Weiland}, {Gold}, {Jarosik}, {Kogut}, {Limon}, {Meyer}, {Tucker}, {Wollack},
  \& {Wright}}]{Hinshaw13}
{Hinshaw}, G., {Larson}, D., {Komatsu}, E., {et~al.} 2013, \apjs, 208, 19,
  \dodoi{10.1088/0067-0049/208/2/19}

\bibitem[{{Hoaglin} {et~al.}(1983){Hoaglin}, {Mosteller}, \&
  {Tukey}}]{Hoaglin83}
{Hoaglin}, D.~C., {Mosteller}, F., \& {Tukey}, J.~W. 1983, {Understanding
  robust and exploratory data anlysis}

\bibitem[{{Ilbert} {et~al.}(2006){Ilbert}, {Arnouts}, {McCracken},
  {Bolzonella}, {Bertin}, {Le F{\`e}vre}, {Mellier}, {Zamorani}, {Pell{\`o}},
  {Iovino}, {Tresse}, {Le Brun}, {Bottini}, {Garilli}, {Maccagni}, {Picat},
  {Scaramella}, {Scodeggio}, {Vettolani}, {Zanichelli}, {Adami}, {Bardelli},
  {Cappi}, {Charlot}, {Ciliegi}, {Contini}, {Cucciati}, {Foucaud}, {Franzetti},
  {Gavignaud}, {Guzzo}, {Marano}, {Marinoni}, {Mazure}, {Meneux}, {Merighi},
  {Paltani}, {Pollo}, {Pozzetti}, {Radovich}, {Zucca}, {Bondi}, {Bongiorno},
  {Busarello}, {de La Torre}, {Gregorini}, {Lamareille}, {Mathez}, {Merluzzi},
  {Ripepi}, {Rizzo}, \& {Vergani}}]{Ilbert06}
{Ilbert}, O., {Arnouts}, S., {McCracken}, H.~J., {et~al.} 2006, \aap, 457, 841,
  \dodoi{10.1051/0004-6361:20065138}

\bibitem[{{Ilbert} {et~al.}(2013){Ilbert}, {McCracken}, {Le F{\`e}vre},
  {Capak}, {Dunlop}, {Karim}, {Renzini}, {Caputi}, {Boissier}, {Arnouts},
  {Aussel}, {Comparat}, {Guo}, {Hudelot}, {Kartaltepe}, {Kneib}, {Krogager},
  {Le Floc'h}, {Lilly}, {Mellier}, {Milvang-Jensen}, {Moutard}, {Onodera},
  {Richard}, {Salvato}, {Sanders}, {Scoville}, {Silverman}, {Taniguchi},
  {Tasca}, {Thomas}, {Toft}, {Tresse}, {Vergani}, {Wolk}, \& {Zirm}}]{Ilbert13}
{Ilbert}, O., {McCracken}, H.~J., {Le F{\`e}vre}, O., {et~al.} 2013, \aap, 556,
  A55, \dodoi{10.1051/0004-6361/201321100}

\bibitem[{{Ito} {et~al.}(2020){Ito}, {Kashikawa}, {Toshikawa}, {Overzier},
  {Kubo}, {Uchiyama}, {Liang}, {Onoue}, {Tanaka}, {Komiyama}, {Lee}, {Lin},
  {Marinello}, {Martin}, \& {Shibuya}}]{Ito20}
{Ito}, K., {Kashikawa}, N., {Toshikawa}, J., {et~al.} 2020, \apj, 899, 5,
  \dodoi{10.3847/1538-4357/aba269}

\bibitem[{{Johnston} {et~al.}(2015){Johnston}, {Vaccari}, {Jarvis}, {Smith},
  {Giovannoli}, {H{\"a}u{\ss}ler}, \& {Prescott}}]{Johnston15}
{Johnston}, R., {Vaccari}, M., {Jarvis}, M., {et~al.} 2015, \mnras, 453, 2540,
  \dodoi{10.1093/mnras/stv1715}

\bibitem[{{Kauffmann} {et~al.}(2004){Kauffmann}, {White}, {Heckman},
  {M{\'e}nard}, {Brinchmann}, {Charlot}, {Tremonti}, \&
  {Brinkmann}}]{Kauffmann04}
{Kauffmann}, G., {White}, S. D.~M., {Heckman}, T.~M., {et~al.} 2004, \mnras,
  353, 713, \dodoi{10.1111/j.1365-2966.2004.08117.x}

\bibitem[{{Kauffmann} {et~al.}(2003){Kauffmann}, {Heckman}, {White}, {Charlot},
  {Tremonti}, {Brinchmann}, {Bruzual}, {Peng}, {Seibert}, {Bernardi},
  {Blanton}, {Brinkmann}, {Castander}, {Cs{\'a}bai}, {Fukugita}, {Ivezic},
  {Munn}, {Nichol}, {Padmanabhan}, {Thakar}, {Weinberg}, \&
  {York}}]{Kauffmann03}
{Kauffmann}, G., {Heckman}, T.~M., {White}, S. D.~M., {et~al.} 2003, \mnras,
  341, 33, \dodoi{10.1046/j.1365-8711.2003.06291.x}

\bibitem[{{Kim} {et~al.}(2002){Kim}, {Kepner}, {Postman}, {Strauss}, {Bahcall},
  {Gunn}, {Lupton}, {Annis}, {Nichol}, {Castander}, {Brinkmann}, {Brunner},
  {Connolly}, {Csabai}, {Hindsley}, {Ivezi{\'c}}, {Vogeley}, \& {York}}]{Kim02}
{Kim}, R. S.~J., {Kepner}, J.~V., {Postman}, M., {et~al.} 2002, \aj, 123, 20,
  \dodoi{10.1086/324727}

\bibitem[{{Koyama} {et~al.}(2013){Koyama}, {Smail}, {Kurk}, {Geach}, {Sobral},
  {Kodama}, {Nakata}, {Swinbank}, {Best}, {Hayashi}, \& {Tadaki}}]{Koyama13}
{Koyama}, Y., {Smail}, I., {Kurk}, J., {et~al.} 2013, \mnras, 434, 423,
  \dodoi{10.1093/mnras/stt1035}

\bibitem[{{Kriek} {et~al.}(2006){Kriek}, {van Dokkum}, {Franx}, {Quadri},
  {Gawiser}, {Herrera}, {Illingworth}, {Labb{\'e}}, {Lira}, {Marchesini},
  {Rix}, {Rudnick}, {Taylor}, {Toft}, {Urry}, \& {Wuyts}}]{Kriek06}
{Kriek}, M., {van Dokkum}, P.~G., {Franx}, M., {et~al.} 2006, \apjl, 649, L71,
  \dodoi{10.1086/508371}

\bibitem[{{Kubo} {et~al.}(2013){Kubo}, {Uchimoto}, {Yamada}, {Kajisawa},
  {Ichikawa}, {Matsuda}, {Akiyama}, {Hayashino}, {Konishi}, {Nishimura},
  {Omata}, {Suzuki}, {Tanaka}, {Yoshikawa}, {Alexander}, {Fazio}, {Huang}, \&
  {Lehmer}}]{Kubo13}
{Kubo}, M., {Uchimoto}, Y.~K., {Yamada}, T., {et~al.} 2013, \apj, 778, 170,
  \dodoi{10.1088/0004-637X/778/2/170}

\bibitem[{{Labb{\'e}} {et~al.}(2005){Labb{\'e}}, {Huang}, {Franx}, {Rudnick},
  {Barmby}, {Daddi}, {van Dokkum}, {Fazio}, {Schreiber}, {Moorwood}, {Rix},
  {R{\"o}ttgering}, {Trujillo}, \& {van der Werf}}]{Labbe05}
{Labb{\'e}}, I., {Huang}, J., {Franx}, M., {et~al.} 2005, \apjl, 624, L81,
  \dodoi{10.1086/430700}

\bibitem[{{Laigle} {et~al.}(2016){Laigle}, {McCracken}, {Ilbert}, {Hsieh},
  {Davidzon}, {Capak}, {Hasinger}, {Silverman}, {Pichon}, {Coupon}, {Aussel},
  {Le Borgne}, {Caputi}, {Cassata}, {Chang}, {Civano}, {Dunlop}, {Fynbo},
  {Kartaltepe}, {Koekemoer}, {Le F{\`e}vre}, {Le Floc'h}, {Leauthaud}, {Lilly},
  {Lin}, {Marchesi}, {Milvang-Jensen}, {Salvato}, {Sanders}, {Scoville},
  {Smolcic}, {Stockmann}, {Taniguchi}, {Tasca}, {Toft}, {Vaccari}, \&
  {Zabl}}]{Laigle16}
{Laigle}, C., {McCracken}, H.~J., {Ilbert}, O., {et~al.} 2016, \apjs, 224, 24,
  \dodoi{10.3847/0067-0049/224/2/24}

\bibitem[{{Lee} {et~al.}(2014){Lee}, {Dey}, {Hong}, {Reddy}, {Wilson},
  {Jannuzi}, {Inami}, \& {Gonzalez}}]{Lee14}
{Lee}, K.-S., {Dey}, A., {Hong}, S., {et~al.} 2014, \apj, 796, 126,
  \dodoi{10.1088/0004-637X/796/2/126}

\bibitem[{{Lemaux} {et~al.}(2014){Lemaux}, {Cucciati}, {Tasca}, {Le F{\`e}vre},
  {Zamorani}, {Cassata}, {Garilli}, {Le Brun}, {Maccagni}, {Pentericci},
  {Thomas}, {Vanzella}, {Zucca}, {Amor{\'\i}n}, {Bardelli}, {Capak},
  {Cassar{\`a}}, {Castellano}, {Cimatti}, {Cuby}, {de la Torre}, {Durkalec},
  {Fontana}, {Giavalisco}, {Grazian}, {Hathi}, {Ilbert}, {Moreau}, {Paltani},
  {Ribeiro}, {Salvato}, {Schaerer}, {Scodeggio}, {Sommariva}, {Talia},
  {Taniguchi}, {Tresse}, {Vergani}, {Wang}, {Charlot}, {Contini}, {Fotopoulou},
  {Gal}, {Kocevski}, {L{\'o}pez-Sanjuan}, {Lubin}, {Mellier}, {Sadibekova}, \&
  {Scoville}}]{Lemaux14}
{Lemaux}, B.~C., {Cucciati}, O., {Tasca}, L.~A.~M., {et~al.} 2014, \aap, 572,
  A41, \dodoi{10.1051/0004-6361/201423828}

\bibitem[{{Lemaux} {et~al.}(2018){Lemaux}, {Le F{\`e}vre}, {Cucciati},
  {Ribeiro}, {Tasca}, {Zamorani}, {Ilbert}, {Thomas}, {Bardelli}, {Cassata},
  {Hathi}, {Pforr}, {Smol{\v{c}}i{\'c}}, {Delvecchio}, {Novak}, {Berta},
  {McCracken}, {Koekemoer}, {Amor{\'\i}n}, {Garilli}, {Maccagni}, {Schaerer},
  \& {Zucca}}]{Lemaux18}
{Lemaux}, B.~C., {Le F{\`e}vre}, O., {Cucciati}, O., {et~al.} 2018, \aap, 615,
  A77, \dodoi{10.1051/0004-6361/201730870}

\bibitem[{{Lemaux} {et~al.}(2020){Lemaux}, {Cucciati}, {Le F{\`e}vre},
  {Zamorani}, {Lubin}, {Hathi}, {Ilbert}, {Pelliccia}, {Amor{\'\i}n},
  {Bardelli}, {Cassata}, {Gal}, {Garilli}, {Guaita}, {Giavalisco}, {Hung},
  {Koekemoer}, {Maccagni}, {Pentericci}, {Ribeiro}, {Schaerer}, {Shen},
  {Talia}, {Tomczak}, {Vanzella}, {Vergani}, \& {Zucca}}]{Lemaux20}
{Lemaux}, B.~C., {Cucciati}, O., {Le F{\`e}vre}, O., {et~al.} 2020, arXiv
  e-prints, arXiv:2009.03324.
\newblock \doarXiv{2009.03324}

\bibitem[{{Li} {et~al.}(2008){Li}, {Mo}, \& {Gao}}]{Li08}
{Li}, Y., {Mo}, H.~J., \& {Gao}, L. 2008, \mnras, 389, 1419,
  \dodoi{10.1111/j.1365-2966.2008.13667.x}

\bibitem[{{Malkan} {et~al.}(2017){Malkan}, {Cohen}, {Maruyama}, {Kashikawa},
  {Ly}, {Ishikawa}, {Shimasaku}, {Hayashi}, \& {Motohara}}]{Malkan17}
{Malkan}, M.~A., {Cohen}, D.~P., {Maruyama}, M., {et~al.} 2017, \apj, 850, 5,
  \dodoi{10.3847/1538-4357/aa9331}

\bibitem[{{Mart{\'\i}n-Navarro} {et~al.}(2018){Mart{\'\i}n-Navarro},
  {Vazdekis}, {Falc{\'o}n-Barroso}, {La Barbera}, {Y{\i}ld{\i}r{\i}m}, \& {van
  de Ven}}]{Mart18}
{Mart{\'\i}n-Navarro}, I., {Vazdekis}, A., {Falc{\'o}n-Barroso}, J., {et~al.}
  2018, \mnras, 475, 3700, \dodoi{10.1093/mnras/stx3346}

\bibitem[{{Matsuda} {et~al.}(2005){Matsuda}, {Yamada}, {Hayashino}, {Tamura},
  {Yamauchi}, {Murayama}, {Nagao}, {Ohta}, {Okamura}, {Ouchi}, {Shimasaku},
  {Shioya}, \& {Taniguchi}}]{Matsuda05}
{Matsuda}, Y., {Yamada}, T., {Hayashino}, T., {et~al.} 2005, \apjl, 634, L125,
  \dodoi{10.1086/499071}

\bibitem[{{Merlin} {et~al.}(2015){Merlin}, {Fontana}, {Ferguson}, {Dunlop},
  {Elbaz}, {Bourne}, {Bruce}, {Buitrago}, {Castellano}, {Schreiber}, {Grazian},
  {McLure}, {Okumura}, {Shu}, {Wang}, {Amor{\'\i}n}, {Boutsia}, {Cappelluti},
  {Comastri}, {Derriere}, {Faber}, \& {Santini}}]{Merlin15}
{Merlin}, E., {Fontana}, A., {Ferguson}, H.~C., {et~al.} 2015, \aap, 582, A15,
  \dodoi{10.1051/0004-6361/201526471}

\bibitem[{{Merlin} {et~al.}(2016){Merlin}, {Bourne}, {Castellano}, {Ferguson},
  {Wang}, {Derriere}, {Dunlop}, {Elbaz}, \& {Fontana}}]{Merlin16}
{Merlin}, E., {Bourne}, N., {Castellano}, M., {et~al.} 2016, \aap, 595, A97,
  \dodoi{10.1051/0004-6361/201628751}

\bibitem[{{Miller} {et~al.}(2018){Miller}, {Chapman}, {Aravena}, {Ashby},
  {Hayward}, {Vieira}, {Wei{\ss}}, {Babul}, {B{\'e}thermin}, {Bradford},
  {Brodwin}, {Carlstrom}, {Chen}, {Cunningham}, {De Breuck}, {Gonzalez},
  {Greve}, {Harnett}, {Hezaveh}, {Lacaille}, {Litke}, {Ma}, {Malkan},
  {Marrone}, {Morningstar}, {Murphy}, {Narayanan}, {Pass}, {Perry}, {Phadke},
  {Rennehan}, {Rotermund}, {Simpson}, {Spilker}, {Sreevani}, {Stark},
  {Strandet}, \& {Strom}}]{Miller18}
{Miller}, T.~B., {Chapman}, S.~C., {Aravena}, M., {et~al.} 2018, \nat, 556,
  469, \dodoi{10.1038/s41586-018-0025-2}

\bibitem[{{Muldrew} {et~al.}(2015){Muldrew}, {Hatch}, \& {Cooke}}]{Muldrew15}
{Muldrew}, S.~I., {Hatch}, N.~A., \& {Cooke}, E.~A. 2015, \mnras, 452, 2528,
  \dodoi{10.1093/mnras/stv1449}

\bibitem[{{Muldrew} {et~al.}(2018){Muldrew}, {Hatch}, \& {Cooke}}]{Muldrew18}
---. 2018, \mnras, 473, 2335, \dodoi{10.1093/mnras/stx2454}

\bibitem[{{Muzzin} {et~al.}(2013){Muzzin}, {Marchesini}, {Stefanon}, {Franx},
  {McCracken}, {Milvang-Jensen}, {Dunlop}, {Fynbo}, {Brammer}, {Labb{\'e}}, \&
  {van Dokkum}}]{Muzzin13}
{Muzzin}, A., {Marchesini}, D., {Stefanon}, M., {et~al.} 2013, \apj, 777, 18,
  \dodoi{10.1088/0004-637X/777/1/18}

\bibitem[{{Noll} {et~al.}(2009){Noll}, {Burgarella}, {Giovannoli}, {Buat},
  {Marcillac}, \& {Mu{\~n}oz-Mateos}}]{Noll09}
{Noll}, S., {Burgarella}, D., {Giovannoli}, E., {et~al.} 2009, \aap, 507, 1793,
  \dodoi{10.1051/0004-6361/200912497}

\bibitem[{{Oke} \& {Gunn}(1983)}]{Oke83}
{Oke}, J.~B., \& {Gunn}, J.~E. 1983, \apj, 266, 713, \dodoi{10.1086/160817}

\bibitem[{{Overzier}(2016)}]{Overzier16}
{Overzier}, R.~A. 2016, \aapr, 24, 14, \dodoi{10.1007/s00159-016-0100-3}

\bibitem[{{Overzier} {et~al.}(2008){Overzier}, {Bouwens}, {Cross}, {Venemans},
  {Miley}, {Zirm}, {Ben{\'\i}tez}, {Blakeslee}, {Coe}, {Demarco}, {Ford},
  {Homeier}, {Illingworth}, {Kurk}, {Martel}, {Mei}, {Oliveira},
  {R{\"o}ttgering}, {Tsvetanov}, \& {Zheng}}]{Overzier08}
{Overzier}, R.~A., {Bouwens}, R.~J., {Cross}, N.~J.~G., {et~al.} 2008, \apj,
  673, 143, \dodoi{10.1086/524342}

\bibitem[{{Postman} {et~al.}(2005){Postman}, {Franx}, {Cross}, {Holden},
  {Ford}, {Illingworth}, {Goto}, {Demarco}, {Rosati}, {Blakeslee}, {Tran},
  {Ben{\'\i}tez}, {Clampin}, {Hartig}, {Homeier}, {Ardila}, {Bartko},
  {Bouwens}, {Bradley}, {Broadhurst}, {Brown}, {Burrows}, {Cheng}, {Feldman},
  {Golimowski}, {Gronwall}, {Infante}, {Kimble}, {Krist}, {Lesser}, {Martel},
  {Mei}, {Menanteau}, {Meurer}, {Miley}, {Motta}, {Sirianni}, {Sparks}, {Tran},
  {Tsvetanov}, {White}, \& {Zheng}}]{Postman05}
{Postman}, M., {Franx}, M., {Cross}, N.~J.~G., {et~al.} 2005, \apj, 623, 721,
  \dodoi{10.1086/428881}

\bibitem[{{Pozzetti} {et~al.}(2010){Pozzetti}, {Bolzonella}, {Zucca},
  {Zamorani}, {Lilly}, {Renzini}, {Moresco}, {Mignoli}, {Cassata}, {Tasca},
  {Lamareille}, {Maier}, {Meneux}, {Halliday}, {Oesch}, {Vergani}, {Caputi},
  {Kova{\v{c}}}, {Cimatti}, {Cucciati}, {Iovino}, {Peng}, {Carollo}, {Contini},
  {Kneib}, {Le F{\'e}vre}, {Mainieri}, {Scodeggio}, {Bardelli}, {Bongiorno},
  {Coppa}, {de la Torre}, {de Ravel}, {Franzetti}, {Garilli}, {Kampczyk},
  {Knobel}, {Le Borgne}, {Le Brun}, {Pell{\`o}}, {Perez Montero},
  {Ricciardelli}, {Silverman}, {Tanaka}, {Tresse}, {Abbas}, {Bottini}, {Cappi},
  {Guzzo}, {Koekemoer}, {Leauthaud}, {Maccagni}, {Marinoni}, {McCracken},
  {Memeo}, {Porciani}, {Scaramella}, {Scarlata}, \& {Scoville}}]{Pozzetti10}
{Pozzetti}, L., {Bolzonella}, M., {Zucca}, E., {et~al.} 2010, \aap, 523, A13,
  \dodoi{10.1051/0004-6361/200913020}

\bibitem[{{Ramella} {et~al.}(2001){Ramella}, {Boschin}, {Fadda}, \&
  {Nonino}}]{Ramella01}
{Ramella}, M., {Boschin}, W., {Fadda}, D., \& {Nonino}, M. 2001, \aap, 368,
  776, \dodoi{10.1051/0004-6361:20010071}

\bibitem[{Richardson(1972)}]{Richardson:72}
Richardson, W.~H. 1972, J. Opt. Soc. Am., 62, 55,
  \dodoi{10.1364/JOSA.62.000055}

\bibitem[{{Santini} {et~al.}(2019){Santini}, {Merlin}, {Fontana}, {Magnelli},
  {Paris}, {Castellano}, {Grazian}, {Pentericci}, {Pilo}, \&
  {Torelli}}]{Santini19}
{Santini}, P., {Merlin}, E., {Fontana}, A., {et~al.} 2019, \mnras, 486, 560,
  \dodoi{10.1093/mnras/stz801}

\bibitem[{{Schenker} {et~al.}(2013){Schenker}, {Ellis}, {Konidaris}, \&
  {Stark}}]{Schenker13}
{Schenker}, M.~A., {Ellis}, R.~S., {Konidaris}, N.~P., \& {Stark}, D.~P. 2013,
  \apj, 777, 67, \dodoi{10.1088/0004-637X/777/1/67}

\bibitem[{{Shi} {et~al.}(2020){Shi}, {Toshikawa}, {Cai}, {Lee}, \&
  {Fang}}]{Shi20}
{Shi}, K., {Toshikawa}, J., {Cai}, Z., {Lee}, K.-S., \& {Fang}, T. 2020, \apj,
  899, 79, \dodoi{10.3847/1538-4357/aba626}

\bibitem[{{Shi} {et~al.}(2019{\natexlab{a}}){Shi}, {Lee}, {Dey}, {Huang},
  {Malavasi}, {Hung}, {Inami}, {Ashby}, {Duncan}, {Xue}, {Reddy}, {Hong},
  {Jannuzi}, {Cooper}, {Gonzalez}, {R{\"o}ttgering}, {Best}, \&
  {Tasse}}]{Shi191}
{Shi}, K., {Lee}, K.-S., {Dey}, A., {et~al.} 2019{\natexlab{a}}, \apj, 871, 83,
  \dodoi{10.3847/1538-4357/aaf85d}

\bibitem[{{Shi} {et~al.}(2019{\natexlab{b}}){Shi}, {Huang}, {Lee}, {Toshikawa},
  {Bowen}, {Malavasi}, {Lemaux}, {Cucciati}, {Le Fevre}, \& {Dey}}]{Shi192}
{Shi}, K., {Huang}, Y., {Lee}, K.-S., {et~al.} 2019{\natexlab{b}}, \apj, 879,
  9, \dodoi{10.3847/1538-4357/ab2118}

\bibitem[{{Shimakawa} {et~al.}(2018){Shimakawa}, {Kodama}, {Hayashi},
  {Prochaska}, {Tanaka}, {Cai}, {Suzuki}, {Tadaki}, \& {Koyama}}]{Shimakawa18}
{Shimakawa}, R., {Kodama}, T., {Hayashi}, M., {et~al.} 2018, \mnras, 473, 1977,
  \dodoi{10.1093/mnras/stx2494}

\bibitem[{{Snyder} {et~al.}(2012){Snyder}, {Brodwin}, {Mancone}, {Zeimann},
  {Stanford}, {Gonzalez}, {Stern}, {Eisenhardt}, {Brown}, {Dey}, {Jannuzi}, \&
  {Perlmutter}}]{Snyder12}
{Snyder}, G.~F., {Brodwin}, M., {Mancone}, C.~M., {et~al.} 2012, \apj, 756,
  114, \dodoi{10.1088/0004-637X/756/2/114}

\bibitem[{{Soares-Santos} {et~al.}(2011){Soares-Santos}, {de Carvalho},
  {Annis}, {Gal}, {La Barbera}, {Lopes}, {Wechsler}, {Busha}, \&
  {Gerke}}]{Soares11}
{Soares-Santos}, M., {de Carvalho}, R.~R., {Annis}, J., {et~al.} 2011, \apj,
  727, 45, \dodoi{10.1088/0004-637X/727/1/45}

\bibitem[{{Speagle} {et~al.}(2014){Speagle}, {Steinhardt}, {Capak}, \&
  {Silverman}}]{Speagle14}
{Speagle}, J.~S., {Steinhardt}, C.~L., {Capak}, P.~L., \& {Silverman}, J.~D.
  2014, \apjs, 214, 15, \dodoi{10.1088/0067-0049/214/2/15}

\bibitem[{{Springel} {et~al.}(2005){Springel}, {White}, {Jenkins}, {Frenk},
  {Yoshida}, {Gao}, {Navarro}, {Thacker}, {Croton}, {Helly}, {Peacock}, {Cole},
  {Thomas}, {Couchman}, {Evrard}, {Colberg}, \& {Pearce}}]{Springel05}
{Springel}, V., {White}, S. D.~M., {Jenkins}, A., {et~al.} 2005, \nat, 435,
  629, \dodoi{10.1038/nature03597}

\bibitem[{{Stalin} {et~al.}(2010){Stalin}, {Petitjean}, {Srianand}, {Fox},
  {Coppolani}, \& {Schwope}}]{Stalin10}
{Stalin}, C.~S., {Petitjean}, P., {Srianand}, R., {et~al.} 2010, \mnras, 401,
  294, \dodoi{10.1111/j.1365-2966.2009.15636.x}

\bibitem[{{Stanford} {et~al.}(1998){Stanford}, {Eisenhardt}, \&
  {Dickinson}}]{Stanford98}
{Stanford}, S.~A., {Eisenhardt}, P.~R., \& {Dickinson}, M. 1998, \apj, 492,
  461, \dodoi{10.1086/305050}

\bibitem[{{Steidel} {et~al.}(2005){Steidel}, {Adelberger}, {Shapley}, {Erb},
  {Reddy}, \& {Pettini}}]{Steidel05}
{Steidel}, C.~C., {Adelberger}, K.~L., {Shapley}, A.~E., {et~al.} 2005, \apj,
  626, 44, \dodoi{10.1086/429989}

\bibitem[{{Stott} {et~al.}(2009){Stott}, {Pimbblet}, {Edge}, {Smith}, \&
  {Wardlow}}]{Stott09}
{Stott}, J.~P., {Pimbblet}, K.~A., {Edge}, A.~C., {Smith}, G.~P., \& {Wardlow},
  J.~L. 2009, \mnras, 394, 2098, \dodoi{10.1111/j.1365-2966.2009.14477.x}

\bibitem[{{Thomas} {et~al.}(2005){Thomas}, {Maraston}, {Bender}, \& {Mendes de
  Oliveira}}]{Thomas05}
{Thomas}, D., {Maraston}, C., {Bender}, R., \& {Mendes de Oliveira}, C. 2005,
  \apj, 621, 673, \dodoi{10.1086/426932}

\bibitem[{{Toshikawa} {et~al.}(2014){Toshikawa}, {Kashikawa}, {Overzier},
  {Shibuya}, {Ishikawa}, {Ota}, {Shimasaku}, {Tanaka}, {Hayashi}, {Niino}, \&
  {Onoue}}]{Toshikawa14}
{Toshikawa}, J., {Kashikawa}, N., {Overzier}, R., {et~al.} 2014, \apj, 792, 15,
  \dodoi{10.1088/0004-637X/792/1/15}

\bibitem[{{Toshikawa} {et~al.}(2016){Toshikawa}, {Kashikawa}, {Overzier},
  {Malkan}, {Furusawa}, {Ishikawa}, {Onoue}, {Ota}, {Tanaka}, {Niino}, \&
  {Uchiyama}}]{Toshikawa16}
---. 2016, \apj, 826, 114, \dodoi{10.3847/0004-637X/826/2/114}

\bibitem[{{Tran} {et~al.}(2010){Tran}, {Papovich}, {Saintonge}, {Brodwin},
  {Dunlop}, {Farrah}, {Finkelstein}, {Finkelstein}, {Lotz}, {McLure},
  {Momcheva}, \& {Willmer}}]{Tran10}
{Tran}, K.-V.~H., {Papovich}, C., {Saintonge}, A., {et~al.} 2010, \apjl, 719,
  L126, \dodoi{10.1088/2041-8205/719/2/L126}

\bibitem[{{Trenti} \& {Stiavelli}(2008)}]{Trenti08}
{Trenti}, M., \& {Stiavelli}, M. 2008, \apj, 676, 767, \dodoi{10.1086/528674}

\bibitem[{{Umehata} {et~al.}(2018){Umehata}, {Hatsukade}, {Smail}, {Alexand
  er}, {Ivison}, {Matsuda}, {Tamura}, {Kohno}, {Kato}, {Hayatsu}, {Kubo}, \&
  {Ikarashi}}]{Umehata18}
{Umehata}, H., {Hatsukade}, B., {Smail}, I., {et~al.} 2018, \pasj, 70, 65,
  \dodoi{10.1093/pasj/psy065}

\bibitem[{{van Dokkum} {et~al.}(2003){van Dokkum}, {F{\"o}rster Schreiber},
  {Franx}, {Daddi}, {Illingworth}, {Labb{\'e}}, {Moorwood}, {Rix},
  {R{\"o}ttgering}, {Rudnick}, {van der Wel}, {van der Werf}, \& {van
  Starkenburg}}]{van03}
{van Dokkum}, P.~G., {F{\"o}rster Schreiber}, N.~M., {Franx}, M., {et~al.}
  2003, \apjl, 587, L83, \dodoi{10.1086/375156}

\bibitem[{{Visvanathan} \& {Sandage}(1977)}]{Visvanathan77}
{Visvanathan}, N., \& {Sandage}, A. 1977, \apj, 216, 214,
  \dodoi{10.1086/155464}

\bibitem[{{Wang} {et~al.}(2019){Wang}, {Schreiber}, {Elbaz}, {Yoshimura},
  {Kohno}, {Shu}, {Yamaguchi}, {Pannella}, {Franco}, {Huang}, {Lim}, \&
  {Wang}}]{Wang19}
{Wang}, T., {Schreiber}, C., {Elbaz}, D., {et~al.} 2019, \nat, 572, 211,
  \dodoi{10.1038/s41586-019-1452-4}

\bibitem[{{Wechsler} {et~al.}(2006){Wechsler}, {Zentner}, {Bullock},
  {Kravtsov}, \& {Allgood}}]{Wechsler06}
{Wechsler}, R.~H., {Zentner}, A.~R., {Bullock}, J.~S., {Kravtsov}, A.~V., \&
  {Allgood}, B. 2006, \apj, 652, 71, \dodoi{10.1086/507120}

\bibitem[{{Williams} {et~al.}(2009){Williams}, {Quadri}, {Franx}, {van Dokkum},
  \& {Labb{\'e}}}]{Williams09}
{Williams}, R.~J., {Quadri}, R.~F., {Franx}, M., {van Dokkum}, P., \&
  {Labb{\'e}}, I. 2009, \apj, 691, 1879, \dodoi{10.1088/0004-637X/691/2/1879}

\bibitem[{{Yang} {et~al.}(2010){Yang}, {Zabludoff}, {Eisenstein}, \&
  {Dav{\'e}}}]{Yang10}
{Yang}, Y., {Zabludoff}, A., {Eisenstein}, D., \& {Dav{\'e}}, R. 2010, \apj,
  719, 1654, \dodoi{10.1088/0004-637X/719/2/1654}

\bibitem[{{Yuan} {et~al.}(2019){Yuan}, {Burgarella}, {Corre}, {Buat},
  {Boquien}, \& {Shen}}]{Yuan19}
{Yuan}, F.-T., {Burgarella}, D., {Corre}, D., {et~al.} 2019, \aap, 631, A123,
  \dodoi{10.1051/0004-6361/201935975}

\bibitem[{{Zavala} {et~al.}(2019){Zavala}, {Casey}, {Scoville}, {Champagne},
  {Chiang}, {Dannerbauer}, {Drew}, {Fu}, {Spilker}, {Spitler}, {Tran},
  {Treister}, \& {Toft}}]{Zavala19}
{Zavala}, J.~A., {Casey}, C.~M., {Scoville}, N., {et~al.} 2019, \apj, 887, 183,
  \dodoi{10.3847/1538-4357/ab5302}

\bibitem[{{Zentner} {et~al.}(2014){Zentner}, {Hearin}, \& {van den
  Bosch}}]{Zentner14}
{Zentner}, A.~R., {Hearin}, A.~P., \& {van den Bosch}, F.~C. 2014, \mnras, 443,
  3044, \dodoi{10.1093/mnras/stu1383}

\end{thebibliography}
\bibliographystyle{aasjournal}



\end{document}